\documentclass[]{aa}  

\usepackage{graphicx}

\usepackage{caption}
\usepackage{subcaption}

\usepackage{txfonts}
\usepackage{xcolor}
 \usepackage[bookmarks=true,pdfnewwindow=true,colorlinks=true,citecolor=blue,final=true]{hyperref}
\bibpunct{(}{)}{;}{a}{}{,} 

\begin{document} 


\title{To Grow Old and Peculiar: A Survey of Anomalous Variable Stars in M80 and Age Determination using K2 and Gaia}

\titlerunning{Anomalous Variable Stars in
M80 with K2 and Gaia}
\authorrunning{Moln\'ar et al.}


   \author{L\'aszl\'o Moln\'ar\inst{1,2,3,4}
          \and
           Emese Plachy\inst{1,2,3,4}
          \and
          Attila B\'odi\inst{1,2,3}
          \and
          Andr\'as P\'al\inst{1,2,4}
          \and
          Meridith Joyce\inst{1,2,3,5}
          \and
          Csilla Kalup\inst{1,2,4}
          \and 
          Christian I. Johnson\inst{5}
          \and
          Zoltán Dencs\inst{6,7}
          \and
          Szabolcs M\'esz\'aros\inst{6,7}
          \and
          Henryka Netzel\inst{1,2,6,8}
          \and
          Karen Kinemuchi\inst{9}
          \and
          Juna A.\ Kollmeier\inst{10}
          \and
          Jose Luis Prieto\inst{11,12}
          \and
          Aliz Derekas\inst{6,7,13}
          }

   \institute{Konkoly Observatory, Research Centre for Astronomy and Earth Sciences, E\"otv\"os Lor\'and Research Network (ELKH), Konkoly Thege Mikl\'os \'ut 15-17, H-1121 Budapest, Hungary\\ 
             \email{molnar.laszlo@csfk.org}
             \and
             CSFK, MTA Centre of Excellence, Budapest, Konkoly Thege Miklós út 15-17., H-1121, Hungary 
             \and
             MTA CSFK Lend\"ulet Near-Field Cosmology Research Group, 1121, Budapest, Konkoly Thege Mikl\'os \'ut 15-17, Hungary 
             \and
             ELTE E\"otv\"os Lor\'and University, Institute of Physics and Astronomy, 1117, P\'azm\'any P\'eter s\'et\'any 1/A, Budapest, Hungary 
             \and
             Space Telescope Science Institute, 3700 San Martin Drive, Baltimore, MD 21218, USA 
             \and
             ELTE E\"otv\"os Lor\'and University, Gothard Astrophysical Observatory, Szent Imre h. u. 112, 9700, Szombathely, Hungary 
             \and
             MTA-ELTE Lend{\"u}let "Momentum" Milky Way Research Group, Hungary 
             \and 
             Institute of Physics, \'Ecole Polytechnique F\'ed\'erale de Lausanne (EPFL), Observatoire de Sauverny, 1290 Versoix, Switzerland 
             \and 
             Apache Point Observatory/New Mexico State University, 2001 Apache Point Road, Sunspot, NM 88349, USA
             \and
             Canadian Institute for Theoretical Astrophysics, University of Toronto, 60 St. George Street, Toronto, ON M5S 3H8, Canada 
             \and
             N\'ucleo de Astronom\'ia de la Facultad de Ingenier\'ia y Ciencias, Universidad Diego Portales, Av. Ej\'ercito 441, Santiago, Chile 
             \and
             Millennium Institute of Astrophysics, Santiago, Chile 
             \and 
             ELKH-SZTE Stellar Astrophysics Research Group, H-6500 Baja, Szegedi út, Kt. 766, Hungary} 
   \date{}

 
  \abstract{The globular cluster Messier 80 was monitored by the \textit{Kepler} space telescope for 80 days during the K2 mission. Continuous, high-precision photometry of such an old, compact cluster allows us to study its variable star population in unprecedented detail. We extract light curves for 27 variable stars using differential-image photometry. A search for new variables in the images led to the discovery of two new variable stars: an RR~Lyrae and a variable red giant star, respectively. Analysis of the RR~Lyrae population reveals multiple RRc stars with additional modes and/or peculiar modulation cycles. We newly classify star V28 as a spotted extreme horizontal branch variable. Despite their faintness, we clearly detect the three SX~Phe stars but we did not find new pulsation modes beyond the known ones in them. Spectra taken with the VLT and Magellan Clay telescopes, as well as absolute color-magnitude diagrams of the cluster based on \textit{Gaia} and Pan-STARRS observations confirm the classification of the peculiar modulated variables as bona fide RRc stars. We propose that they highlight a subgroup of overtone stars that may have been overlooked before. We fit MESA isochrones to the CMDs to estimate the age and metallicity of the cluster. We confirm that M80 is old and metal-poor, but show that isochrone fitting to old populations comes with numerous uncertainties.  }


   \maketitle
%

\section{Introduction} \label{sec:intro}
The globular Messier 80 was discovered, among with many other galactic globular clusters, by \citet{Messier-1781}. It lies close to the Galactic bulge, and it is famous for the bright nova eruption of T~Sco in 1860 \citep{Sawyer-1938-nova,Shara-1995,Dieball-2010}. 

\object{M80} (or NGC 6093) is an old, compact, metal-poor globular cluster. Its age is estimated to be roughly between 10.5--13.5 Gyr \citep{salaris-weiss-2002,Marin-Franch-2009}, but high levels of interstellar extinction and difficulties with fitting the main sequence turnoff point made it difficult to narrow the age down further \citep{vandenberg-2013}. 

The metallicity of M80 has been easier to pinpoint through spectroscopic observations of its red giant population. \citet{Carretta-2009} and \citet{Carretta-2015} derived [Fe/H] = $-1.75\pm0.08$ dex and $-1.791\pm0.082$ dex, respectively, while the meta-analysis of \citet{Bailin-2019} found [Fe/H] = $-1.789\pm 0.003$ with an intrinsic spread of $0.014^{+0.003}_{-0.002}$~dex. The distance of the cluster was measured by many authors with various methods: the meta-analysis of \citet{Baumgardrt-2021} found it to be $10.34 \pm 0.12$~kpc, which translates to a parallax and a distance modulus of $\varpi = 0.0967\pm0.0011$~mas and $\mu=15.073\pm0.025$~mag, respectively.

M80 was observed by the \textit{Hubble} space telescope multiple times. The resulting color-magnitude diagrams (CMDs) were fitted with isochrones calculated from three different stellar evolutionary codes by \citet{BarkerPaust2018}. The best-fitting Dartmouth Stellar Evolution Program (DSEP), PAdova and TRieste Stellar Evolution Code (PARSEC), and MESA Isochrones and Stellar Tracks (MIST) models provided progressively lower [Fe/H] indices (--1.65, --1.68 and --1.90) and higher ages (12.5, 13.5 and 14.0 Gyr), respectively. Their model fits used larger distance moduli than usual, between 15.5 and 15.6 magnitudes: while these agreed with the quoted results of \citet{Ruelas-Mayorga-2016}, this result appears to be an outlier and virtually all other estimates are 0.5 mag lower, as evidenced by the meta-study of \citet{Baumgardrt-2021}.

\subsection{Variable stars in M80}

Despite the interest generated by T~Sco, no further variable stars were identified in M80 until \citet{Pickering-Bailey-1898} and \citet{Bailey-1902} reported the first two discoveries, followed by five more by \citet{Sawyer-1942} and \citet{Sawyerhogg-1955}. The first extensive work was published by \citet{Wehlau-1990}, based on decades of photographic data. Their work included ten variables, although two of those, R~Sco and S~Sco, initially labeled as V6 and V7, turned out to be nearby Mira stars that are not associated with the cluster. More recently, \citet{Shara-2005} discovered two more variables, and then \citet{Kopacki-2013} carried out a thorough photometric survey and discovered twenty new variables, thirteen of which are pulsating stars. Light variations were detected in three stars in far-ultraviolet photometry, two of which were known RR~Lyrae stars, while the third turned out to be a new SX~Phe-type pulsator \citep{Thompson-2010}. \citet{Salinas-2016} confirmed previous uncertain variable star identifications with adaptive optics imaging photometry. Finally, \citet{Figuera-2016} identified six further variable stars in the core of the cluster, all of which show slow, low-amplitude brightness changes. 

The two brightest variable stars, which were also identified first, are red giants: M80-V1 is a W Vir-type Cepheid variable, and V2 is a possible semiregular star with uncertain periodicity. Of the two, V1 has been studied more extensively: based on the 16.0 d pulsation period of the star, \citet{Sawyer-1942} estimated a distance of 16 kpc to M80, and \citet{Plachy-2017} identified cycle-to-cycle variations in the star from the continuous photometry obtained during the K2 mission of the \textit{Kepler} space telescope. 

The majority of the other variables in the cluster are pulsating stars, with 15 RR~Lyrae and four SX~Phoenicis stars identified so far. These identified variables have been cataloged by \citet{clement-2001}\footnote{\url{https://www.astro.utoronto.ca/~cclement/read.html}}, and we follow their V--numerical variable notation here. The number of RRab and RRc stars are nearly the same, with seven RRab and eight RRc stars identified, which is typical for Oosterhoff II-type clusters \citep{bono-1994}.

\subsection{Globular clusters with K2}

The \textit{Kepler} space telescope observed multiple globular clusters during the K2 mission \citep{Howell2014,Borucki-2016}. This was the first time that extended high-precision photometry was obtained for these objects: earlier observations with the \textit{Hubble} space telescope, for example, only collected multicolor snapshots to study the color-magnitude distributions of the cluster members instead \citep{Sarajedini-2007,Piotto-2015}. Moreover, earlier space photometry missions were limited in resolution and depth due to smaller telescope apertures, and in the number of target stars due to processing power and data storage constraints. 

The necessity to repurpose \textit{Kepler} into the K2 mission opened up the possibility to observe a wide variety of targets, and to experiment with new applications. Stellar clusters were covered by larger continuous pixel mosaics, including eight globular clusters over the mission (M4, M9, M19, M80, NGC5897, NGC6293, NGC6355 and Terzan 5), even though the limited resolution of the telescope, at 4"/px, was a clear deficit when observing stellar fields as dense as the insides of globular clusters \citep{cody-2018}.   

Before our study, K2 light curves have been studied in depth only for M4, the closest and loosest of these eight clusters \footnote{Results on the oscillating red giants and mass loss in M80 were published after the completion of this paper by \citet{howell-2023}.}. \citet{Miglio-2016} investigated the asteroseismic properties of bright red giants in the cluster, and found excellent agreement with independent distance moduli and stellar masses based on isochrone fitting. \citet{Howell-2022} and \citet{tailo-2022} both estimated the integrated mass loss between the late stages of stellar evolution, based on asteroseismic mass estimates for several cluster members. The light curves of the stellar population of the cluster were searched for variable stars and exoplanets \citep{wallace-m4variables-2019,wallace-m4planets-2020}. Moreover, stars with millimagnitude amplitude variations were identified on the horizontal branch, labeled as ultra-low amplitude RR Lyrae stars \citep{wallace-ularrl-2019}. Additional modes in some overtone RR Lyrae stars were reported by \citet{kuehn-2017}. In contrast, only a single W~Vir-type variable at the outskirts of M80 has been studied so far from all the other clusters that were observed in the K2 mission \citep{Plachy-2017}. 

\section{Data and methods} \label{sec:data}
Our work is largely based on the broad-band visible photometry collected with the \textit{Kepler} space telescope. 
We also analyzed the available astrometric and photometric data from Data Release 3 (EDR3) of the \textit{Gaia} mission \citep{gaia2016,gaia-edr3-2021}, and acquired spectra for some targets.

\begin{figure*}
\begin{center}
\noindent
\resizebox{58mm}{!}{\includegraphics{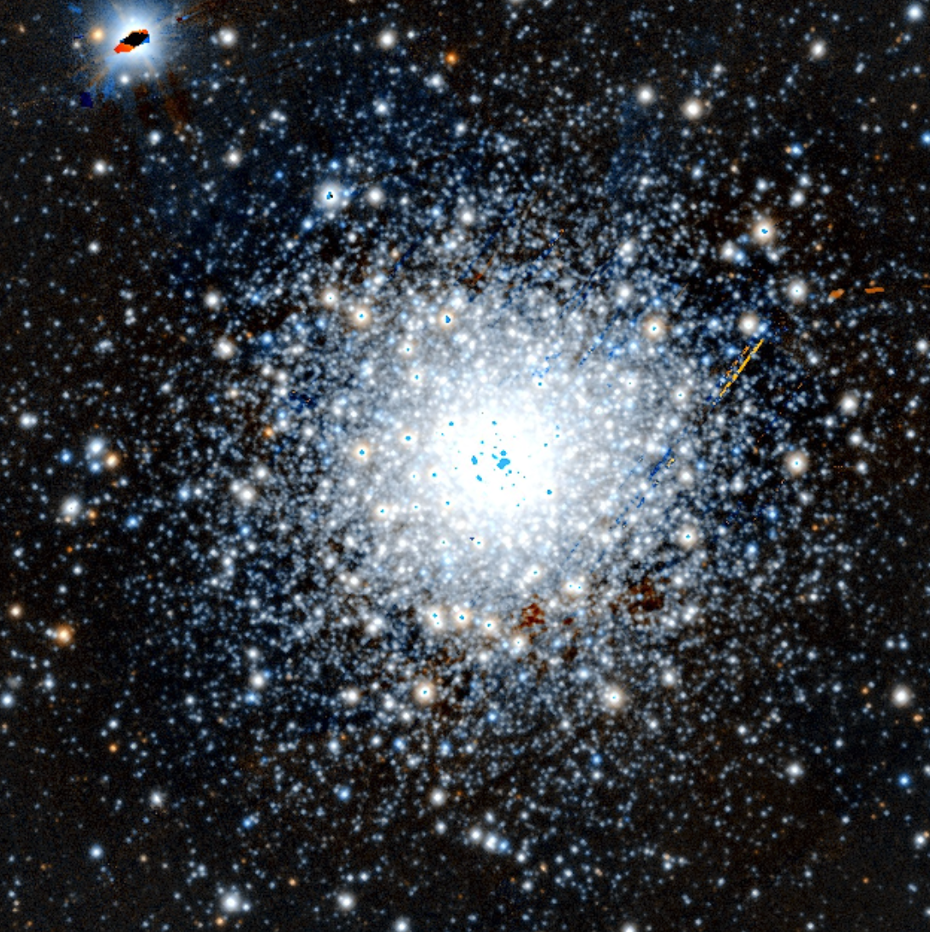}}\hspace*{3mm}%
\resizebox{58mm}{!}{\includegraphics{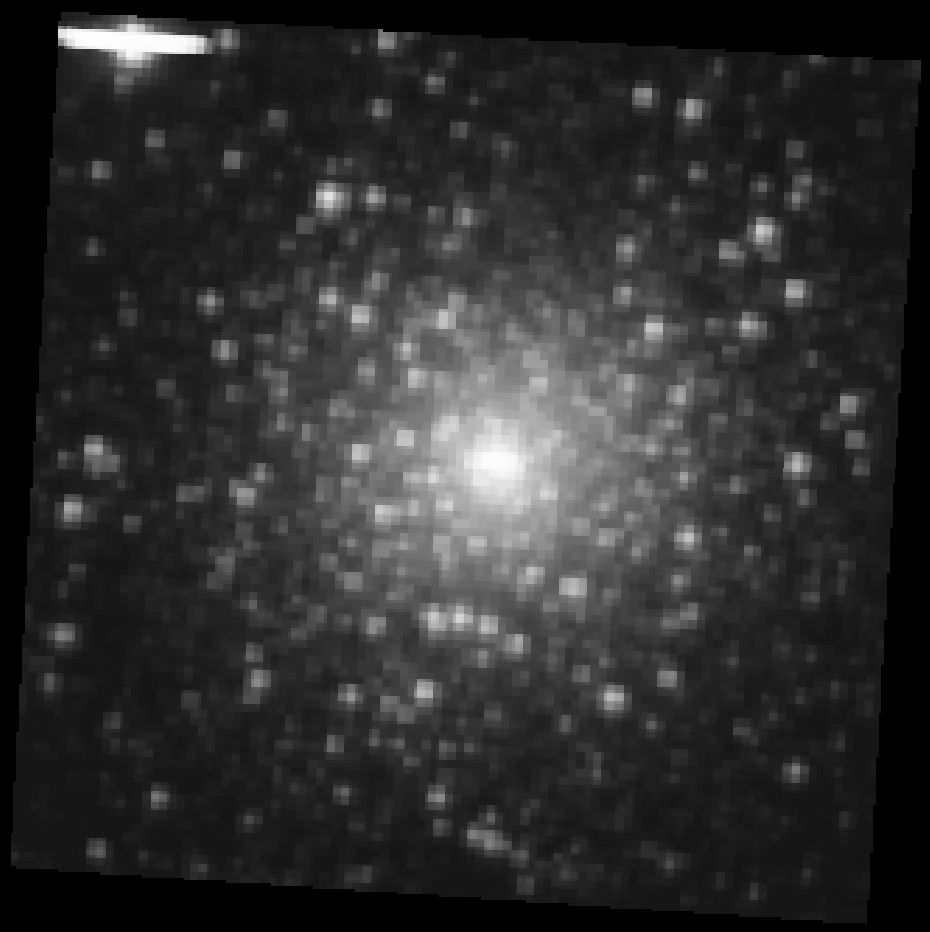}}\hspace*{3mm}%
\resizebox{58mm}{!}{\includegraphics{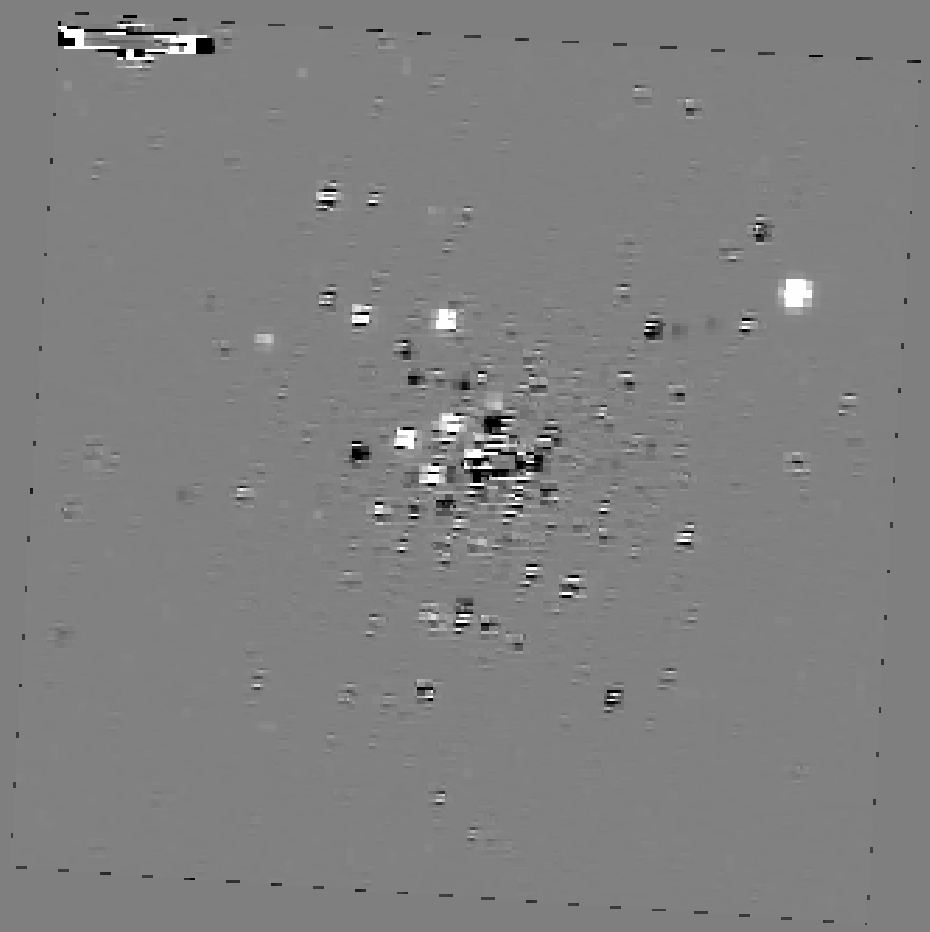}}
\end{center}
\caption{Images of M80, from left to right: Pan-STARRS DR1 color image (\textit{g} and \textit{z} bands; left); the enlarged K2 master image (center); a differential image, showing some of the variables as white or black spots, plus the residual structures in the image (right). Images are aligned to equatorial coordinates, with North being up and East to the left, and are about $430"\times430"$ in size.   }
\label{fig:gc-map}
\end{figure*}

\subsection{K2 photometry}
M80 was observed during Campaign 2 of the K2 mission. This campaign collected data from the vicinity of the Galactic center and the Upper Scorpius star-forming region in late 2014 for 77.1~d. The campaign was rather uneventful, save for brief periods of scattered light caused by Mars passing through the field-of-view of the telescope. A few frames were also discarded at the beginning of the campaign due to poor pointing performance.

M80 was covered with a $100\times100$ px continuous pixel mask, which translates to a field-of-view of $400"\times400"$. Given the compactness of M80, with a half-light radius of only $r_h=37"$, this easily encompasses most of the cluster, even though the Jacobi radius of M80 is considerably larger at $r_J = 20.9'$ (\citealt{harris-1996} (2010 edition); \citealt{deBoer-2019}). Images were collected in long cadence mode, with an integration time of 29.4 min per frame. 

For the data reduction, we used the versatile and tested differential-image photometry method based on the FITSH software package \citep{fitsh}. Most other photometry methods rely on pixel photometry, i.e., summing the flux from a discrete set of pixels and then correcting for image motions afterward. However, pixel photometry can be difficult to achieve for very dense stellar fields that also move about in the images. We, therefore, followed the more conventional method of registering the images to the same reference frame and then applying classical aperture photometry to differential images. This method was used already to collect photometry for a variety of moving objects in the Solar System \citep{farkas-takacs-2017,Molnar-2018,Marton-2020,Szabo-2020,Kalup-2021}, but also for stellar, stationary targets \citep{Plachy-2017,Vida-2017}, and the details have been presented in the works referenced.   

In brief, we collected the K2 target pixel files containing the images of the cluster and a few nearby stars and combined them into a larger image. We then calculated the astrometric solutions for all the frames of the campaign, anchoring the astrometry to the brighter stars on these mosaic images, based on an initial solution for the Full Field Images acquired during the campaign. These steps were necessary to transform the images into the same reference frame and to correct for the attitude changes of the telescope. Since only two reaction wheels were operable during the K2 mission, the telescope was freely rotating about the optical axis due to torque generated by solar radiation pressure. The drifting attitude was reset to the unstable equilibrium point with thrusters every six hours, causing a quasi-periodic instrumental signals in the photometry. 

We then enlarged each frame linearly in both the RA and Dec dimensions by 3.1 times, distributing the flux levels from the original pixels among the new ones. We employ this step to lessen the effects of low pixel resolution and sharp stellar PSFs (point spread functions) which may lead to strong fringing caused by the sub-pixel position differences between the aligned images. This is especially important to ensure smoother images after subtracting the master frame in the next step. Smaller pixels also allow us finer control over the aperture sizes.

With the aligned frames at hand, we created a master frame from a subset of images and then subtracted that from every individual image. Aperture photometry was carried out on the differential images. We then used the \textit{Gaia} EDR3 \textit{G} magnitudes as reference brightness since the two missions use very similar bandpasses, and their magnitude values have been found to correlate well \citep[see, e.g.,][]{molnar-gaia-2018}, and shifted the differential light curves to the calculated median flux levels, using the \textit{Kepler} flux calibration of \citet{Lund2015}. 

One limitation of our method is that it affects the measured amplitudes. We tested apertures between 3--5 px in the enlarged images (the former equalling $\approx 1$ px in the original images) to limit contamination from neighboring stars. While these tight apertures led to cleaner light curves, they come at the cost of cutting off the outer parts of the PSFs and thus underestimating the variation amplitudes. 

\begin{figure*}
\includegraphics[width=1.0\textwidth]{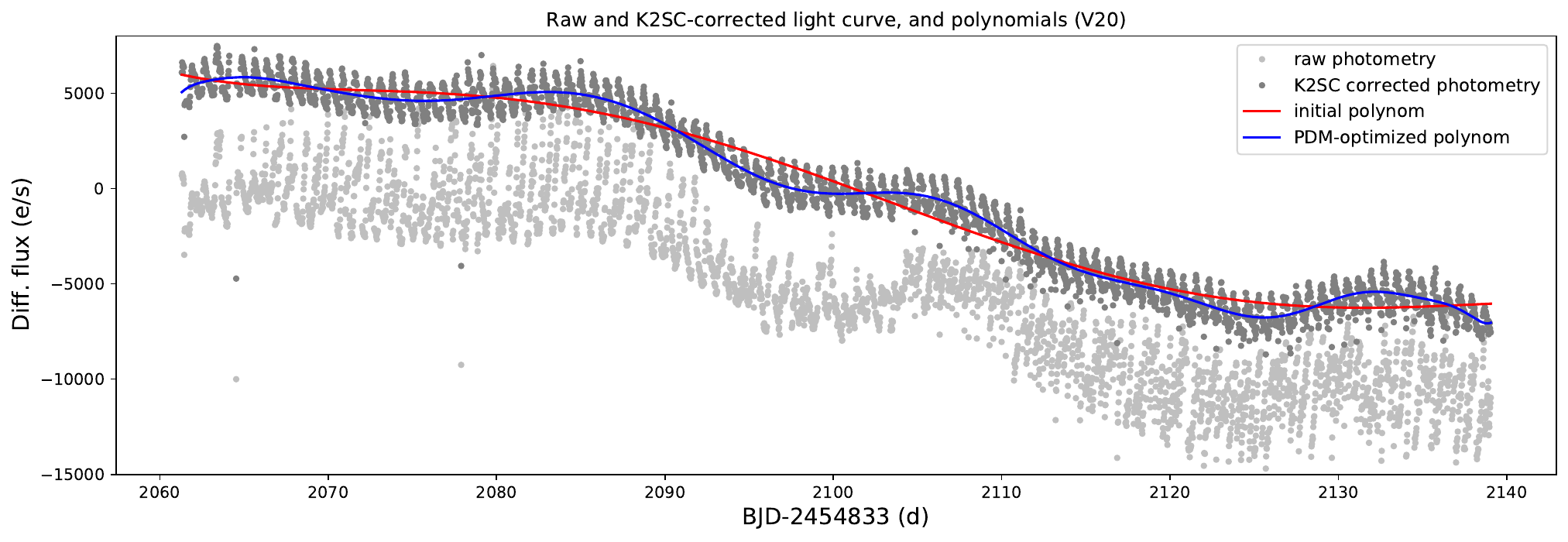}
\includegraphics[width=1.0\textwidth]{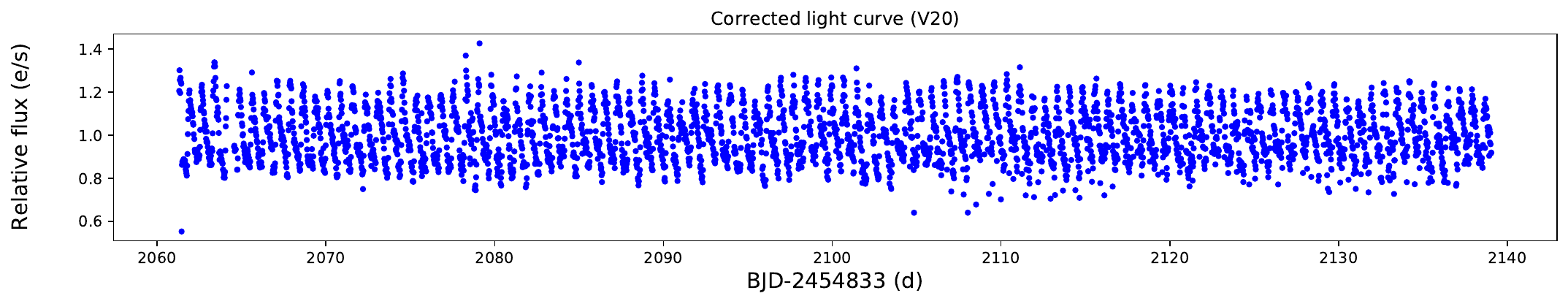}
\includegraphics[width=1.0\textwidth]{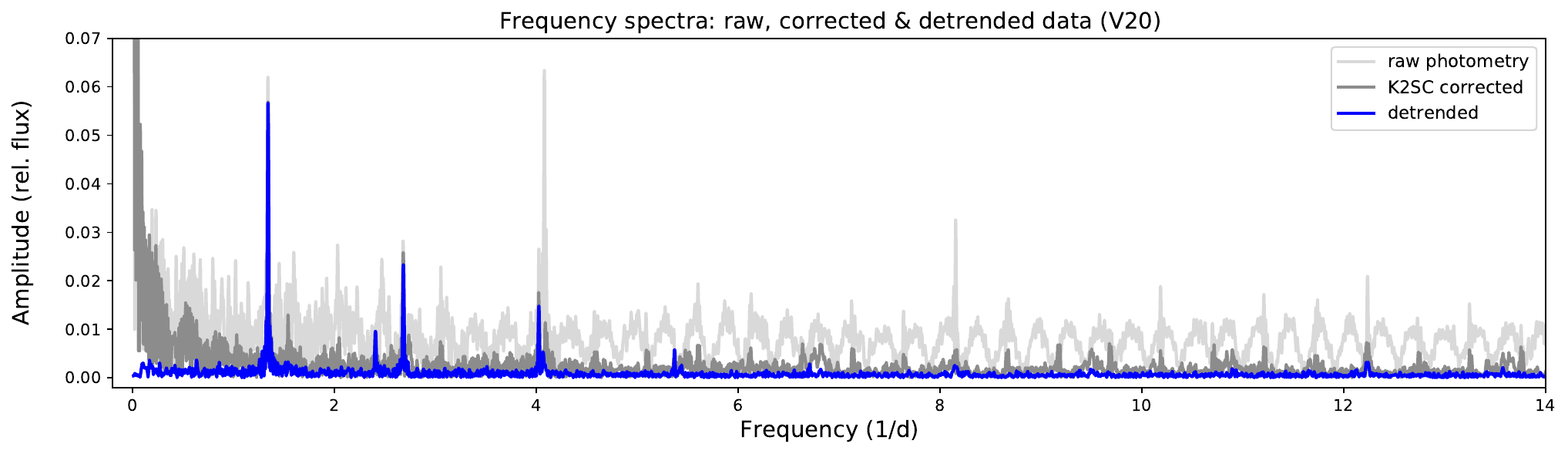}
\caption{Light curve correction steps demonstrated on RR Lyrae star V20, which is close to the core and hence strongly affected by systematics. Top: raw differential-image photometry (light gray) and after we applied the K2SC method (dark gray). The red and blue lines are the initial 20th degree polynomial and the final, optimized with through PDM. Middle: the corrected and detrended light curve, in relative flux. Bottom: frequency spectra of the light curves at each step. }
\label{fig:v20-corr}
\end{figure*}

\subsection{K2 light curve processing}
The very dense stellar field meant that the raw light curves from the aperture photometry often contained various systematics and instrumental signals, especially those close to the core or other bright stars. The main source of these systematics were the attitude changes of the telescope, causing characteristic six-hour changes in the light curves. As the Sun was moving closer to the field-of-view, two further, slower changes occurred: the amount of zodiacal light increased in the background, and the sizes of the stellar images also increased due to temperature-induced focus changes. Therefore, the light curves required further processing to remove these to the best possible extent. 

However, RR Lyrae light curves can be notoriously hard to correct as they include a high-amplitude signal with timescales similar to the frequency of the attitude corrections \citep{Plachy-2021}. One post-processing method we found to work exceptionally well with RR Lyrae light curves is the K2 Systematics Correction algorithm \citep[K2SC,][]{K2SC-2016}. This method uses Gaussian Processes to simultaneously fit two signals in the data, with the intrinsic variation assumed to be (quasi-)periodic in time, whereas the systematics assumed to correlate with the pixel positions of the source. 

While K2SC is an excellent tool to remove the fast systematics related to the attitude changes, it preserves most of the slow trends in the data. This is beneficial if the slow trends are intrinsic to the targets, but the trends present in the RR Lyrae light curves are of instrumental origin. We removed these with a polynomial-fitting algorithm, where we first determined the pulsation period, then optimized the coefficients of the polynomial by minimizing the phase dispersion of the phase-folded light curve \citep{autoeap-2022}. This way we ensured that the pulsation cycles overlap not only in phase but in flux as well, and  the polynomial follows the average brightness of the star much more closely than a simple least-squares fit to the light curve would, even for polynomial orders in the 10-20 range. We present one particularly difficult case in Fig.~\ref{fig:v20-corr}.  

In a few cases, we opted to scale the light curve with the calculated trend instead of subtracting it. Scaling is equivalent to correcting any sensitivity changes in the photometry, such as changes to the PSF size or shape compared to the aperture, which might introduce flux loss. 

The final step of the photometric post-processing was the removal of clear outlier points via sigma clipping the light curve. We then proceeded with the frequency analysis, as described in Sect.~\ref{sect:results}, where we identified cross-contamination between some of the light curves. In the final light curves presented in Table~\ref{tab:lc} in Appendix~\ref{app}, contamination coming from nearby known variable stars have been prewithened, so that the light curves contain variations intrinsic to the targeted variable star to the highest extent possible.

\subsection{Spectroscopy}
We obtained spectra from two telescopes for selected targets. Observations were collected with the LDSS3 (Low Dispersion Survey Spectrograph) instrument installed on the 6.5~m Magellan Clay telescope at Las Campanas Observatory. Five RRc stars (V3, V4, V13, V14, V16) were observed on May 20, 2017. One spectrum per star was recorded with 450 or 600~s exposure times, using the VPH-All grism that covers the whole visible spectrum (425--1000~nm) and has a resolution of R = 860. 

We carried out observations with the Dual Imaging Spectrograph (DIS) on the 3.5~m ARC (Astrophysics Research Consortium) telescope at Apache Point Observarory in New Mexico. Observations were collected on multiple occasions, but were challenging given the low altitude of the cluster and the faintness of the targets. Although the resolution of these spectra were higher than the LDSS3 ones, their SNR turned out to be too low to be useful to us. 

Three stars, V03, V14 and V16 were observed with the GIRAFFE instrument on the VLT U2 (Kueyen) telescope in 2004 as part of a larger program targeting HB stars in the cluster \citep{m80-vlt-2004}. We downloaded the wavelength-calibrated spectra from the ESO archive and analyzed them along with the others. These are high-resolution (R $\sim$ 20,000) but low SNR ($\sim$ 20) spectra, but we were still able to estimate some atmospheric parameters from them.

\section{Variability search}
We searched the \textit{Kepler} pixels for new variables. We visually inspected the differential images and identified two previously unknown variable sources. One star, [CBG2015]~16162 (as identified by \citealt{Carretta-2015}), is a red giant that exhibits slow variations, not unlike V2. This star is cataloged as EPIC~204275713 and Gaia~DR3 6050422207724345216, in the K2 Ecliptic Plane Input Catalog (EPIC) and in Gaia DR3, respectively \citep{EPIC-2016,gaia-edr3-2021}. We clearly detect slow changes in its brightness, but the K2 observations are too short to estimate any periodicity. The other star, EPIC~204270257 (Gaia~DR3 6050421928546399360) is likely an eclipsing binary, with an orbital period of 0.24842~d.  

We also searched the pixels systematically. We binned the image down to 200$\times$200~px then generated the frequency spectra of each 2$\times$2~px aperture of the image, shifting this square by one pixel along each row and then by one row vertically. We computed the Fourier spectra in each position and inspected them visually. We found one previously unknown modulated RRc star, [MMP2009] M80~15470, this way, which was originally cataloged as a horizontal branch (HB) star by \citet{moni-bidin_2009}. 

We also extracted photometry from the K2 images for other \textit{Gaia} DR3 sources found on the HB near the known RR Lyrae stars. Neither of them turned out to be a pulsator. 

\section{Photometric results of RR Lyrae stars}
\label{sect:results}
Prior to our study, 15 RR Lyrae stars were known in M80, of which seven are RRab and eight are RRc stars. We discovered the ninth RRc star in the cluster. For some of them, however, light curve shapes raise doubts about their cluster membership and/or their classification as RR~Lyrae stars. 

\begin{figure*}
\includegraphics[width=1.0\textwidth]{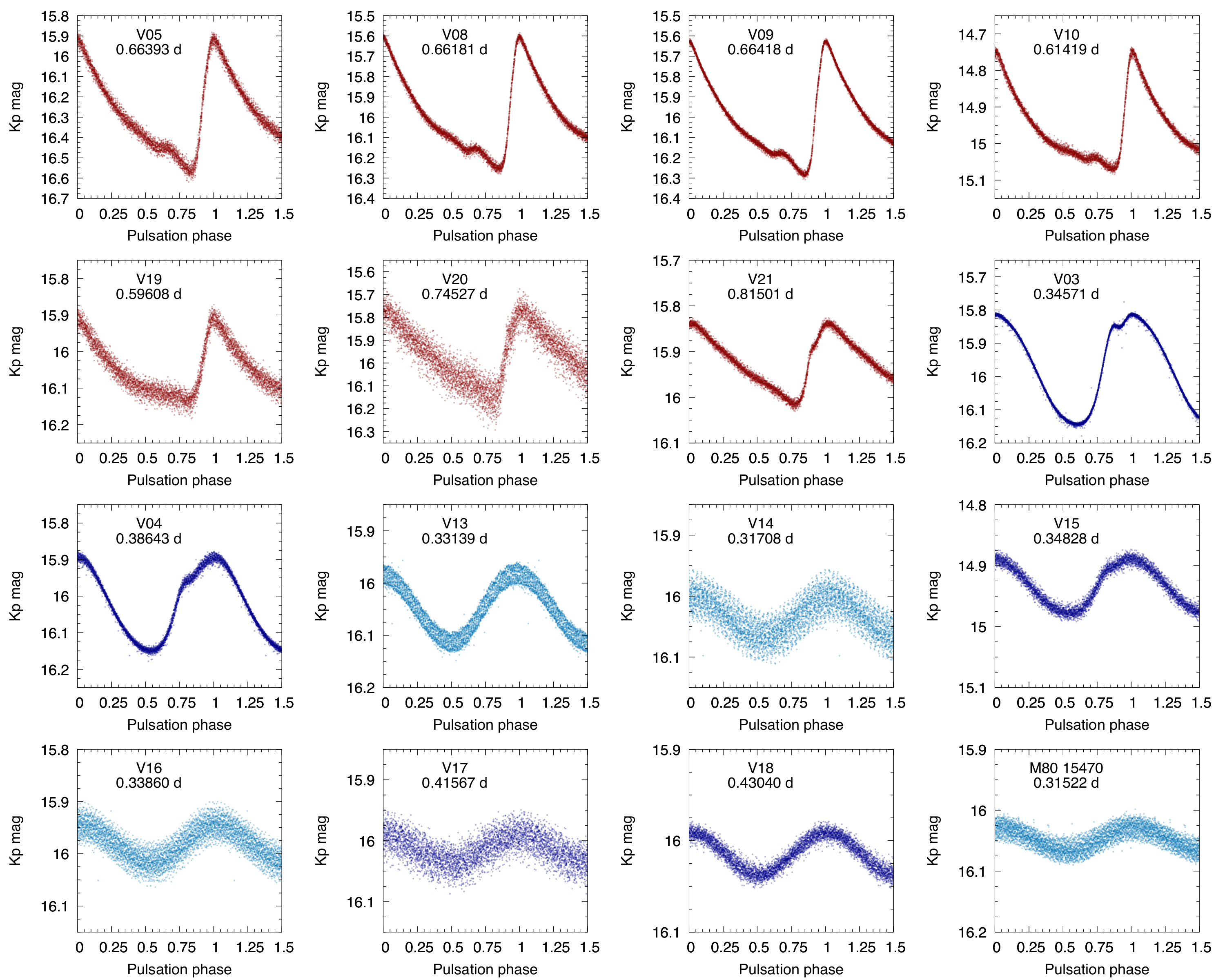}
\caption{Phase curves of the RR Lyrae stars. Corrected K2 light curves are folded with the indicated pulsation periods. Red: RRab stars; dark blue: normal RRc stars; light blue: peculiar RRc stars. }
\label{fig:rrl-folded}
\end{figure*}

We analysed the K2 light curves with the \texttt{Period04} software that calculates multi-frequency fits for the light curves based on their sine-based Fourier transforms \citep{period04}. Phase curves folded with the dominant period are presented in Fig.~\ref{fig:rrl-folded}. As the figure shows, the scatter of the phase curves varies considerably. For some, such as for V03 and V09, the curves are narrow, as expected from the photometric capabilities of \textit{Kepler}. The large scatter in V17, V19 and V20 is due to their positions close to the core of the cluster, where background contamination and systematics are stronger. In contrast, the spread in V13, V14, V16 and M80~15470 is intrinsic, and is caused by modulation. 

We then computed the relative Fourier parameters, the amplitude ratios and phase differences of the first harmonics for each star. These are defined as $R_{i1} = A_i/A_1$ and $\phi_{i1} = \phi_i-i\phi_1$ by \citet{simonlee1981} and then first used for RR~Lyrae stars by \citet{simon-teays-1982}, where $i=2$ and 3 refer to the first and second harmonics of the pulsation frequency, respectively. Our threshold for significant detection was SNR > 4.0 (where SNR is the signal-to-noise ratio). For some stars we were unable to detect even the first or second harmonics, and thus we could only calculate the SNR = 4.0 upper limits for the $R_{i1}$ values. In these cases, the noise levels were calculated from the smoothed residual frequency spectra of pre-whitened light curves. Relative Fourier parameters are widely used to describe the shape of the light curve and to classify stars into variable classes and subclasses. Parameters were already calculated for a larger collection of stars observed in the K2 mission by \citet{molnar-gaia-2018} and \citet{EAP-2019}. We show the values for the M80 stars against the clusters of RRab and RRc stars published by \citet{molnar-gaia-2018} in Fig.~\ref{fig:rrl-fourier}. Fourier amplitudes, phases and relative Fourier parameters are listed in the Appendix in Tables~\ref{tab:amp}, \ref{tab:phase} and \ref{tab:four}.

Stars from M80 fit relatively well into the RRc and RRab groups in Fig.~\ref{fig:rrl-fourier}. RRab stars mostly populate the long-period, Oosterhoff II side of the RRab locus. RRc stars also populate the long-period side, but multiple of them have very low $R_{21}$ and $R_{31}$ values. These points apparently fall below the RRc group of \citet{molnar-gaia-2018}. However, this discrepancy is more likely to reveal a deficit in the general K2 sample than a peculiarity in the M80 group. All targets for the K2 mission had to be proposed in advance, hence RR Lyrae stars were largely selected from the literature and existing catalogs, often from various ground-based surveys \citep{plachy-2016}. Classification schemes for these surveys often did not label very smooth, sine-like light curves that feature little if any humps and light curve asymmetries as RR Lyrae stars, but rather as eclipsing binary or elliptical variables. This raises the possibility that there might be further overtone RR Lyrae stars in the halo and thick disk with smooth and symmetric, sine-like light curves that have not been recognized as RR Lyrae stars yet.

\begin{figure*}
\includegraphics[width=1.0\textwidth]{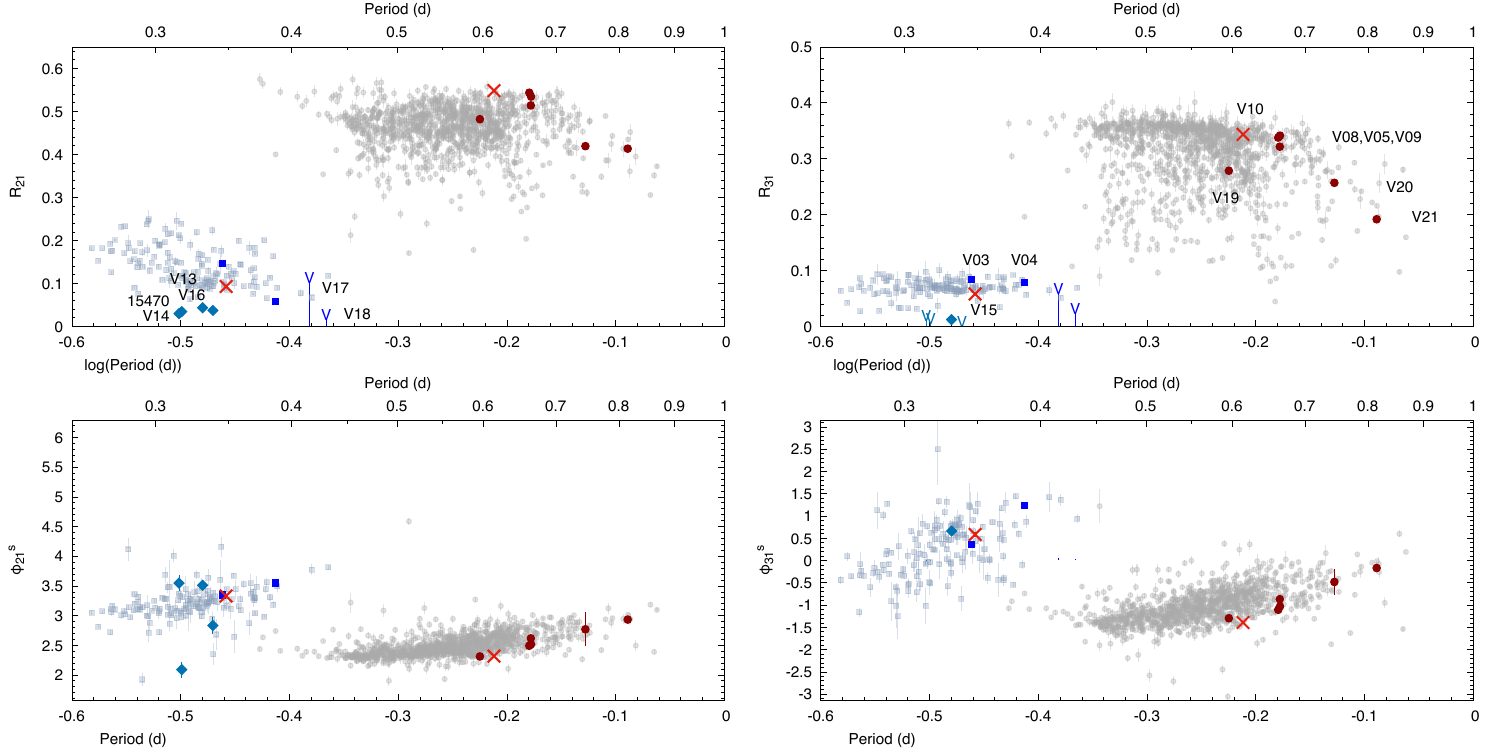}
\caption{Fourier parameters of the RR Lyrae stars. Colors are the same as in Fig.~\ref{fig:rrl-folded}. Red crosses indicate the foreground stars V10 and V15. V signs mark upper limits for the amplitude ratios, where harmonics were not detected. }
\label{fig:rrl-fourier}
\end{figure*}

\subsection{Foreground variables}
Variables V10 and V15 are about 1.1~mag brighter than the rest of the RR~Lyrae stars and the horizontal branch of the cluster, suggesting that they are either not members of M80 or are some other, more luminous variables in the cluster (see Sect.~\ref{sec:cmd} for a detailed discussion of the cluster CMD). Unfortunately, \textit{Gaia} did not acquire useful parallaxes for either of the stars, so we could not decide their membership based on distances alone. The \textit{Gaia} data put both stars above the HB, but the overtone star, V10, which we would expect to be hotter, appears to be the redder of the two. V10 is not present in the Pan-STARRS observations, but V15 is, and it appears to be 2 mag brighter and significantly bluer than the normal RR Lyrae stars.

The relative Fourier parameters of both V10 and V15 agree with the K2 RRab and RRc distributions, respectively, suggesting that they are indeed RR~Lyrae variables. We know from models that He-enhanced RR~Lyrae stars are more luminous than those with canonical abundances, but even a significant excess can only lead to a 0.2--0.5~mag increase in brightness, considerably less than the measured 1.1 mag value \citep{marconi-2018}. Furthermore, we do not expect strong differences in the He content within the population of a globular cluster. Therefore, we conclude that V10 and V15 must be foreground objects.

\subsection{RRab stars}
M80 includes six RRab stars as confirmed members of the cluster. All have rather long periods, between 0.60 and 0.82~d. Two of the stars, V19 and V20, are close to the center of the cluster, which degrades the quality of their K2 photometry. We were able to collect precise photometry for the other four stars. 

We investigated the light curves for signs of modulation and/or extra modes. We identified additional frequencies in multiple stars, but those turned out to be cross-contamination between the pulsators. After we removed these signals from the light curves, we did not detect any signs of modulation or extra modes in the member RRab stars.

We detected a slight amplitude decrease in the assumed foreground star V10, but without any associated phase change. Slow amplitude changes can be caused either by instrumental effects or intrinsic amplitude variation, but in this case its origin remains ambiguous. 

\subsection{RRc stars - unusual light curves and modulation cycles}

\begin{figure*}
\includegraphics[width=1.0\textwidth]{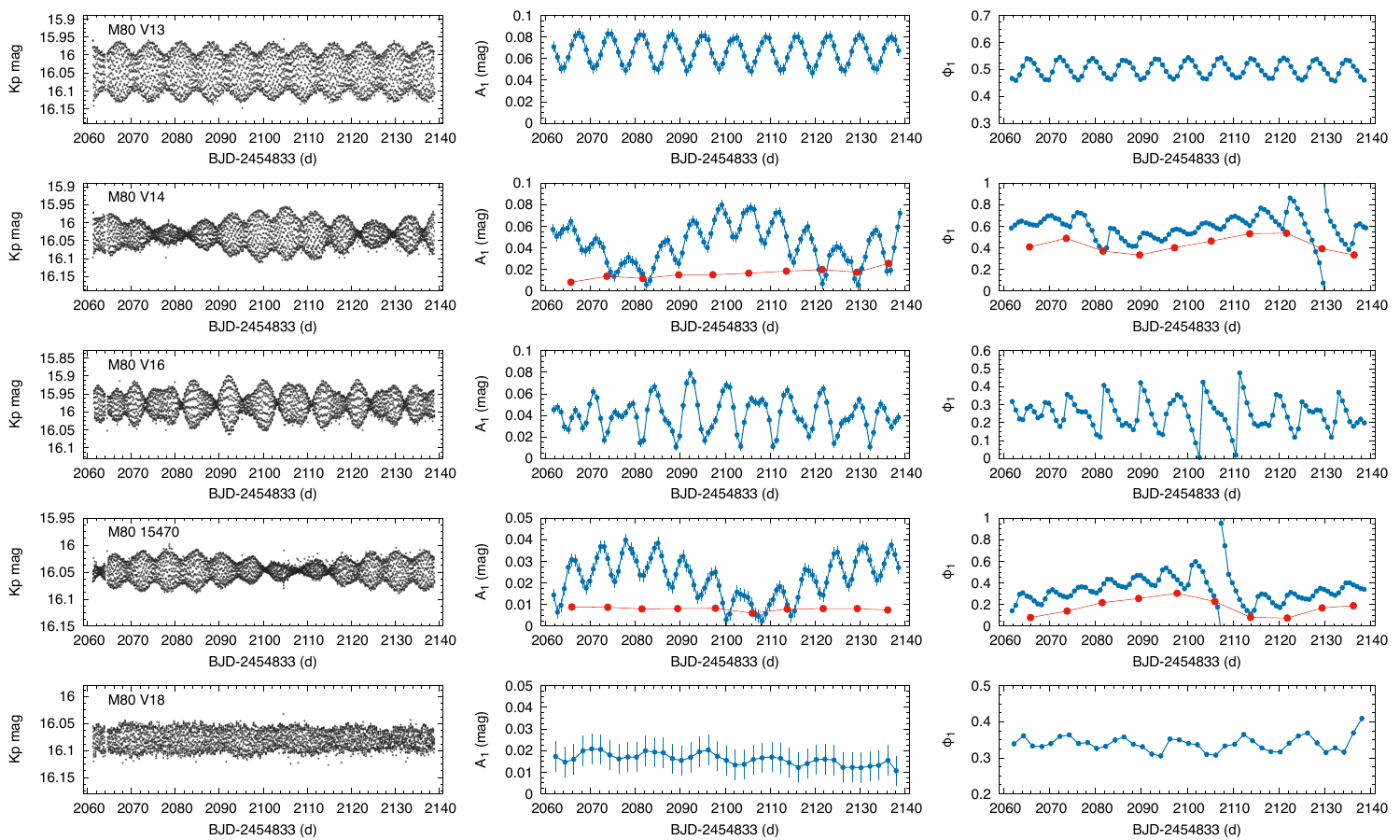}
\caption{The modulated RRc stars. Left column: K2 light curves. Middle column, in blue: variation of $A_1$, the Fourier-amplitude of the pulsation frequency. Right column, in blue: variation of $\phi_1$, the Fourier phase. In red, middle and right: variation of the amplitude and phase of the $A_1$ curve when fitted with the shorter modulation frequency.  }
\label{fig:rrc-bl}
\end{figure*}

\begin{figure*}
\includegraphics[width=1.0\textwidth]{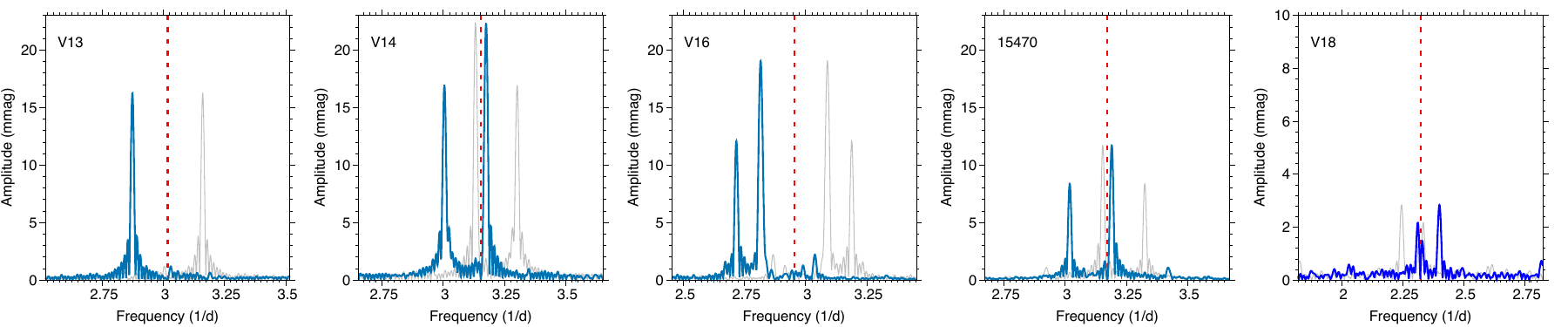}
\caption{Sidepeak structures around the main pulsation frequency (red dashed line), which has been already pre-whitened. The blue line is the residual frequency spectrum, the gray line is the same, but mirrored in frequency around the pulsation peak. Note the lack of overlap among the sidepeaks, and the change in amplitude scale for V18 (dark blue).  }
\label{fig:rrc-side}
\end{figure*}

The most striking result from our photometry was the discovery of multiple peculiar overtone stars with very unusual light curve shapes and strong amplitude and phase modulations. The distribution of peculiar and more regular RRc stars is roughly half and half. Two of the four regular RRc stars, V03 and V04 (as well as the foreground star, V15) appear to be classic overtone stars, with asymmetric light curves and prominent humps ob the rising branches. Two further stars, V17 and V18, have more symmetric light curves with no detectable shockwave features, but this is to be expected from long-period RRc stars, above 0.4~d period. V17 is near the center, and thus its light curve suffers from strong systematics. The light curve of V18 is noisy too, but modulation is clearly detectable in it, especially in the pulsation phase: it is reminiscent of many other Blazhko RRc stars with strong phase modulation. 

The four remaining stars, V13, V14, V16 and the newly discovered M80~15470, however, appear to be drastically different. Each varies with low amplitude and have strongly variable light curves, and they remain almost sinusoidal even at maximum amplitude. To test the modulation properties, we first computed the temporal variation of the pulsation amplitude and phase by cutting the light curve into short segments and fitting the $A_1$ and $\phi_1$ parameters of the same $f_1$ pulsation frequency in each segment independently. The results, along with V18, the other modulated RRc star, are shown in Fig.~\ref{fig:rrc-bl}.

We found that the modulation is strictly periodic in only one of the stars, V13, in which the amplitude variation is moderate as well. In contrast, the modulation in V16 appears to be quasi-periodic, and the pulsation amplitude drops almost to zero in some instances. Finally, in V14 and M80~15470, we find a shorter and a longer modulation cycle operating simultaneously. To test this, we calculated the temporal variations of the fast modulation by segmenting and fitting the $A_1(t)$ data sets in the same way as we did with the light curves (the red points in Fig.~\ref{fig:rrc-bl}). While the amplitude of the fast modulation stayed almost constant in both stars, their phases followed the slow changes present in the $\phi_1(t)$ data. We also detect complete rotations of the pulsation phase in both stars.

We find that the second, longer modulation is present differently in the amplitude and phase variations of V14 and M80~15470. The amplitude modulation appears to be a simple addition of the short and long cycles. But since the phase variations are also added together, the longer cycle can also be detected in the amplitude modulation as a phase modulation of the shorter cycle. In contrast, the phase modulation includes both cycles, but the amplitude of the shorter cycle appears to be also variable, with the cycle jump happening at pulsation amplitude minimum (or amplitude modulation minimum).

The four strongly modulated stars have very asymmetric sidepeak amplitudes. We show the structures of sidepeaks around the first overtone frequency in Fig.~\ref{fig:rrc-side}. We mirrored the  spectra around the position of the main pulsation frequency (in gray) to show how, even if sidepeaks appear both below and above, they do not form symmetric structures, and no significant frequency counterparts can be found that would from a triplet structure with the pulsation frequency. The only sign of a symmetric sidepeak pair can be found in V18, the foreground star, and it corresponds to the slow decrease of the pulsation amplitude on timescales longer than the observations. This can be either a long modulation cycle or a side effect of data processing.  

Stars with a single modulation sidepeak next to the pulsation frequency (also known as doublets) are also called BL1 stars. In the OGLE sample, \citet{netzel-2018} found that 24\% of modulated RRc stars belong to the BL1 group. While it is possible that ground-based observations might not reveal low-amplitude sidepeaks that belong to triplets with very different amplitudes, high-precision space-based observations also revealed several BL1 RRc stars \citep{netzel-2023}. Apparently, the RR Lyrae population of M80 is also peculiar in the way of having only BL1 RRc stars in the cluster. Two very similar stars were discovered by \citet{netzel2019} in the OGLE Bulge data, both having low amplitudes, short periods and two close-by frequency peaks. We discuss the occurences of similar stars further in Section \ref{sect:newclass}.

We determined the modulation periods from the $A_1(t)$ amplitude variation data for each star. This method was considerably more accurate than relying on the sidepeak distances in the frequency spectrum alone for all stars except V13, where results agreed for both methods. In V14 and M80~15470 we were able to determine the periods of the long and short cycles separately. In V16, the modulation appears to be quasi-periodic, and we identified two characteristic timescales based on the frequency spectrum of the amplitude variation. 

\begin{table} 
\centering 
\caption{Modulation periods and period uncertainties of the Blazhko RRc stars.}
\begin{tabular}{lcccc}  
\hline \hline
ID & $P_{\rm mod1}$ (d) & $\sigma P_{\rm mod1}$ (d) & $P_{\rm mod2}$ (d) & $\sigma P_{\rm mod2}$ (d) \\
\hline
V13 & 6.955 & 0.005 & -- & -- \\
V14 & 6.956 & 0.025 & 46.7  & 0.8 \\ 
V16 & 4.244 & 0.013 & 7.303 & 0.025  \\
V18 & 12.90 & 0.10 & -- & --  \\
15470 & 6.503 & 0.023 & 55.3 & 1.0 \\
\hline
\end{tabular}
\end{table}

\begin{figure*}
\includegraphics[width=1.0\textwidth]{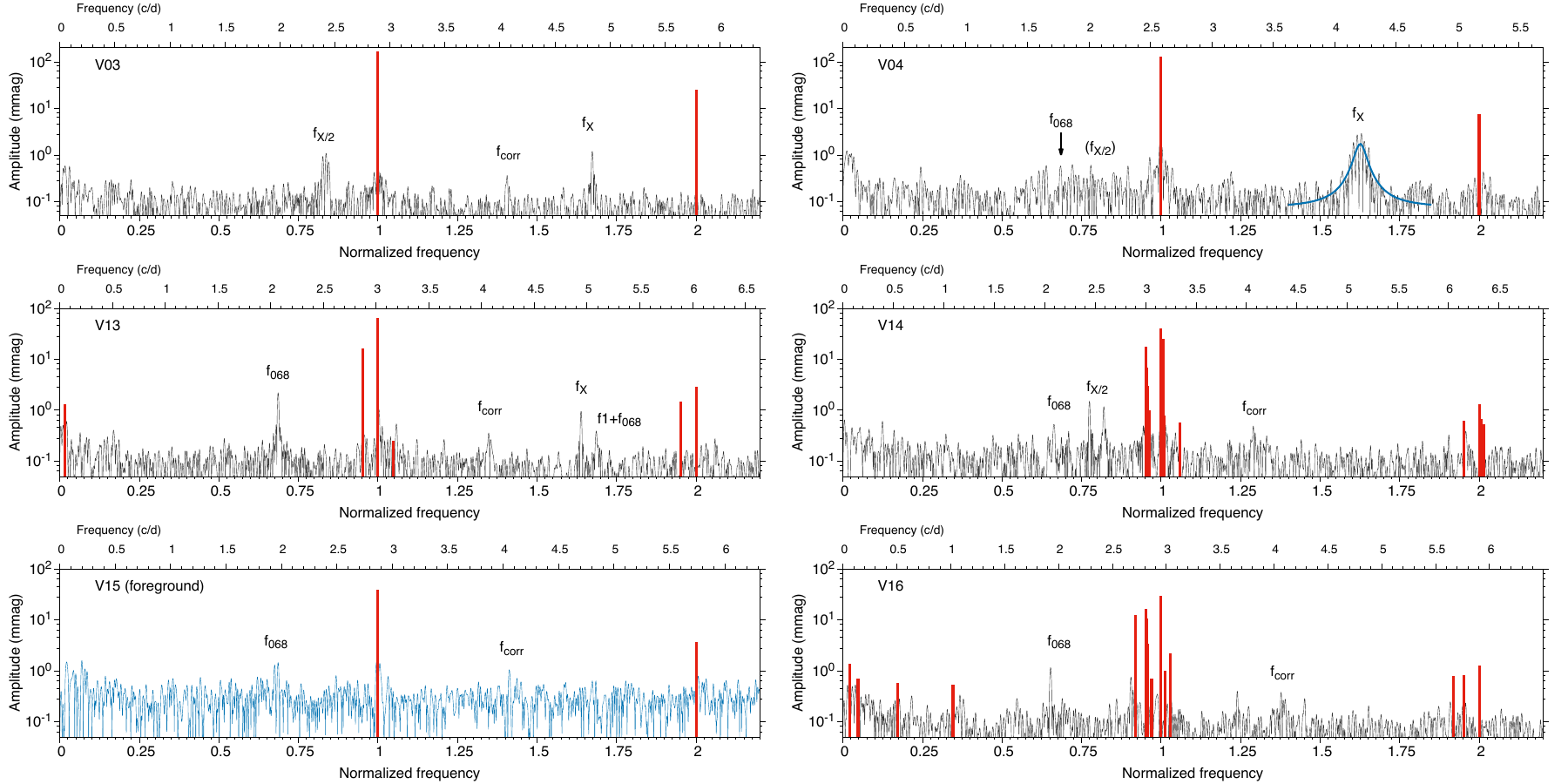}
\caption{Fourier spectra of the RRc stars that exhibit extra modes. Red bars mark frequencies that have been prewhitened. The blue line in V04 shows the Lorentzian fit to the $f_X$ mode.} 
\label{fig:rrc-fspectra}
\end{figure*}

\subsection{Extra modes in RR Lyrae stars}
We now know, thanks to high-precision photometric measurements, that many RR Lyrae stars, and specifically, the majority of RRc stars, are not pure single-mode pulsators. We did not find any extra modes in RRab stars in M80, therefore we focus on the overtone stars here. Additional low-amplitude modes have been identified in large numbers in the OGLE bulge RRc light curves \citep{netzel2015a,netzel2015b,netzel2019}. Similar modes have been observed in a few RRc stars by the space-based CoRoT and Kepler prime missions \citep{szabo2014,moskalik2015,sodor-2017}. Recently, systematic studies of RRc stars observed by TESS and K2 concluded that 81--83\% of first-overtone stars exhibit extra modes \citep{Molnar-2021,Benko-2023}.

\subsubsection{The $f_{0.61}/f_X$-type modes}
Extra modes in RRc stars come in two main varieties. One consists of a collection of short-period modes at $P/P_1$ period ratios, typically between 0.61--0.64. In the bulge, these $P_{0.61}$ or $P_X$ modes form two main ridges and an intermediate group between them in the Petersen diagram. The leading hypothesis for the formation of these modes is that they are high-order, $\ell=8,9$ non-radial modes excited at $f_X/2$ frequencies, but due to geometric effects we often observe the first harmonic peak, $f_X$, with the highest amplitude instead \citep{dziembowski}. The linear model survey of \citet{netzel-2022} showed that these modes are not excited near the blue edge of the instability strip, in agreement with their distributions observed in field stars, indicating that high-$\ell$ modes are a plausible explanation \citep{Molnar-2021,netzel-2023}. Modes with similar period ratios can also be observed in first-overtone classical and anomalous Cepheids \citep{Moskalik-2009,smolec-2016,suveges-2018,Plachy-2021}.

As the lower panel of Fig.~\ref{fig:rrc-fspectra} shows, we detected these signals in four stars (V03, V04, V13, V14), and each case turned out to be rather different. V13 is the simplest case, where we detect a single peak that is very close to the $\ell=9$ ridge. 

In V03, we detect three $f_X$ peaks at the bottom of the period ratio range. For one of those, the middle peak, we also detect the exact $f_X/2$ subharmonic. Notably, the peaks in V03 fall clearly outside of the distribution of the OGLE points, but agree with some of the space-based results. If we compare the positions of these points with the theoretical results of \citet{dziembowski}, we find that the peaks fall into the vicinity of the $\ell=10$ sequence that has virtually no counterparts in the OGLE data, but it is suspected in field stars \citep{Benko-2023}. 

\begin{figure}
\includegraphics[width=1.0\columnwidth]{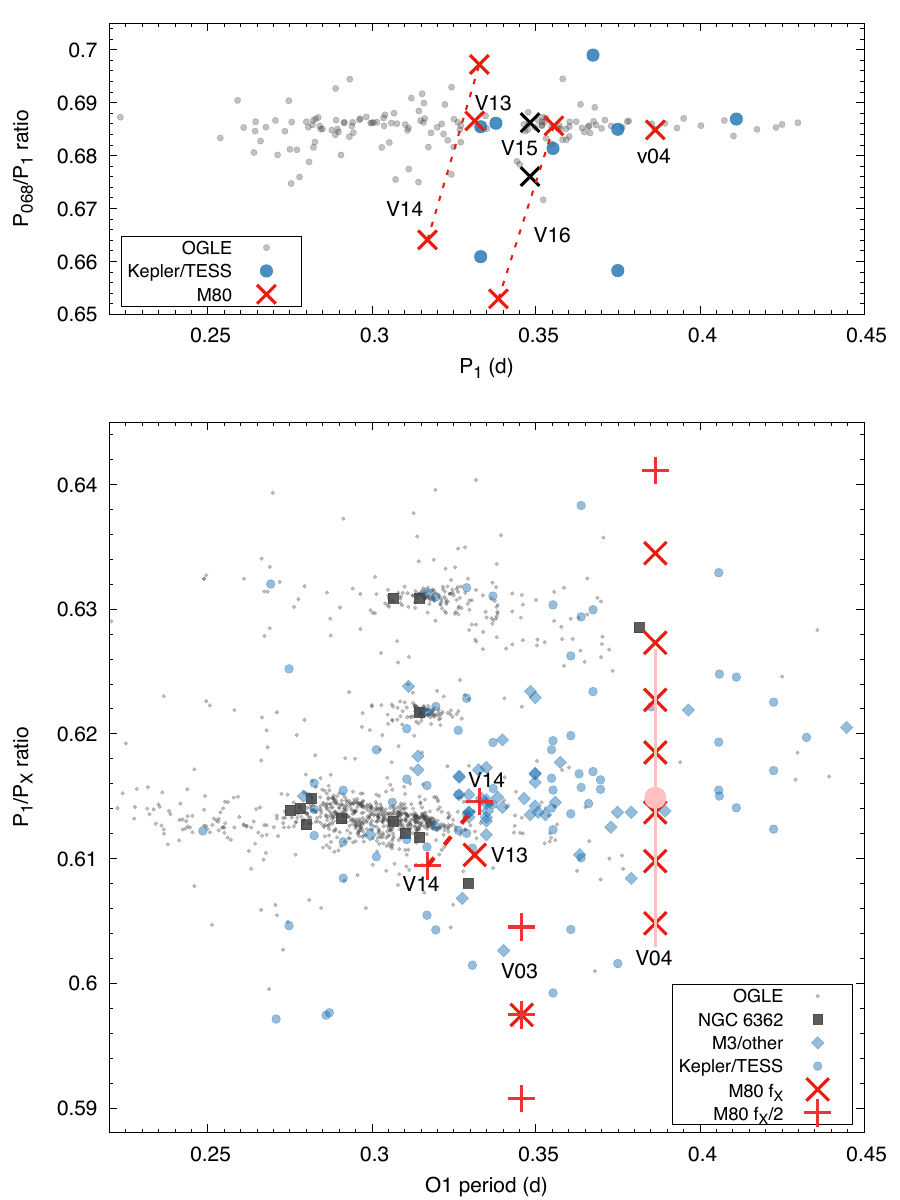}
\caption{Petersen diagrams of the two types of common additional modes in RRc stars. Upper panel: the $f_{0.68}$-type modes. Given the multitude of frequency peaks in V14 and V16, we include ratios between two possible sets of peaks. Lower panel: the $f_{0.61}$- or $f_X$-type modes. Here we plot the regular $f_X$ modes with X crosses, whereas + signs mark the positions of the $f_X/2$ peaks when doubled. For V04, we also show the peak and full width at half maximum of the Lorentzian fit to the peaks with the pink point and line.} Note that in V03 the respective peaks overlap indicating that we observe the half of the $f_X$ peak, but in V04, the $f_X/2$ signal lies outside of a pattern of $f_X$ frequencies. Here, we again show two possible ratios for V14. Underlying points are from \citet{Molnar-2021} and references therein.  
\label{fig:rrc-petersen}
\end{figure}

In V04 we observe a forest of peaks that spread across the diagram, between period ratios 0.605--0.634. Fitting a Lorentzian function to this feature gives us a peak frequency of $4.2085\pm0.0005\,d^{-1}$, corresponding to a period ratio of 0.6149, with a full width at half maximum of 0.1010$\,d^{-1}$. We also detect some amplitude excess around the subharmonic range, but can only detect a single significant peak that would correspond to a signal outside the forest of peaks. If we compare this pattern to the underlying space-based data (blue circles in Fig~\ref{fig:rrc-petersen}), we can see that they also spread across the diagram more evenly than the OGLE sample, but not to the extent as the identified frequencies in V04. This forest of peaks in V04 suggests that the $f_X$ mode (or modes) is either not coherent or strongly variable. 

Finally, in V14, we find no significant peaks above $f_1$, only below that. V14 is one of the modulated stars, and the two peaks repeat the same frequency separation as the two main components, $f_1$ and $f_1-f_m$. The identical frequency separation suggests that in this star, both modes---the first overtone and the extra mode---experience the same modulation. Here, we identify these two peaks as $f_X/2$ subharmonics: even though we do not observe any harmonics in the data.

\subsubsection{The $f_{0.68}$-type modes}
The other type of extra mode occurs less frequently, but it is even more mysterious. This mode appears at periods longer than the first overtone, around the \textit{frequency} ratio $f/f_1 = 0.686$. These are usually labeled as $f_{0.68}$, but despite the unfortunately very similar nomenclature, they are very different from the $f_{0.61}$ modes. One noticeable property of these modes is that they fall below the expected frequency of the fundamental mode, which would be at $f_0/f_1\approx 0.74$. At the moment, we do not have a good hypothesis for the origin of this mode.

This mode is less frequent than the $f_X$ modes in the bulge RR~Lyrae population, but in M80 we identified it in four stars, equalling the number of stars with $f_X$ modes, and it is also present in V15. In V04 and V13, the frequency peaks fall right into the sequence mapped by the OGLE data, whereas V14 is offset from the main ridge. In the suspected foreground star, V15, we found two nearby peaks, one of which is offset from the mode ridge.

V13 is a rare example of an RRc star that is modulated with both types of extra modes being excited. In V14, we detect a low-amplitude peak that is likely a $f_{0.68}$, and two more that appear to be the subharmonics of the $f_X$ peak. The identifications are presented in Fig.~\ref{fig:rrc-fspectra}. 

We can also plot the period ratios of these signals onto a Petersen diagram to examine their distribution more closely. We compare the identified peaks to that of the OGLE sample and the various other discoveries from ground- and space-based observations in Fig.~\ref{fig:rrc-petersen}. For V14 and V16, the strong modulation makes is ambiguous which frequency is the true first overtone, especially if the modulation is caused my beating of nearby modes. Therefore, we plotted frequency ratios belonging to the two highest peaks there. For the $f_{0.68}$ modes either frequency gives plausible ratios, especially since low ratios below 0.67 have been observed in other stars, although in only two so far. For the $f_X$ modes in V14, we only show the pairings of the lower and higher frequency pairings from the $f_X$ and O1 doublet peaks: other combinations would fall outside the period ratio range we show. 

We list the detected extra mode signals for the RRc stars in Table~\ref{tab:rrc-modes}.

\begin{table} 
\centering 
\caption{The detected extra modes in RRc stars. For V04, we only list the two highest peaks from the $f_X$ region.}
\begin{tabular}{lccc}  
\hline \hline
Star & Frequency (d$^{-1}$) & Amplitude (mmag) & Mode \\
\hline 
V03 & 2.3926(6) & 0.90(8) & $f_{\rm X}/2$ \\
V03 & 2.4208(5) & 1.06(8) & $f_{\rm X}/2$ \\
V03 & 2.4483(8) & 0.70(9) & $f_{\rm X}/2$ \\
V03 & 4.8416(5) & 1.15(8) & $f_{\rm X}$ \\
\hline
V04 & 1.772(2) & 0.60(15) & $f_{0.68}$ \\
V04 & 1.867(2) & 0.62(15) & ? \\
V04 & 2.018(2) & 0.53(15) & $f_{\rm X}/2$ \\
V04 & 4.1840(4) & 2.53(11) & $f_{\rm X}$ \\
V04 & 4.2168(4) & 2.88(11) & $f_{\rm X}$ \\
\hline
V13 & 2.0716(3) & 2.17(9) & $f_{0.68}$ \\
V13 & 4.9446(6) & 0.97(9) & $f_{\rm X}$ \\
\hline
V14 & 2.4446(6) & 1.50(11) & $f_{\rm X}/2$ \\
V14 & 2.5883(8) & 1.17(11) & $f_{\rm X}/2$ \\
V14 & 2.095(2) & 0.51(15) & $f_{0.68}$ \\
\hline
V16 & 1.9282(7) & 1.12(10) & $f_{0.68}$ \\
\hline
\end{tabular}
\label{tab:rrc-modes}
\end{table}

\subsubsection{Mode detection limits}
We identified extra modes in 5 out of 8 member RRc stars (62.5\%) and in zero RRab stars. These incidence rates are clearly lower than the $82\%$ and 35\% rates found in brighter field RRc and RRab stars \citep{Molnar-2021,Benko-2023}. This could indicate either an intrinsically lower presence of extra modes in the cluster members, or a difference in detection sensitivity between the two data sets. To test this, we calculated the upper detection limits for the stars where no extra modes were found. This limit was set again as the SNR = 4.0 level at the position of either the expected $f_X$ mode $(1.62\,f_1)$ in RRc stars or at $1.5\,f_0$ in RRab stars. 

\begin{figure}
\includegraphics[width=1.0\columnwidth]{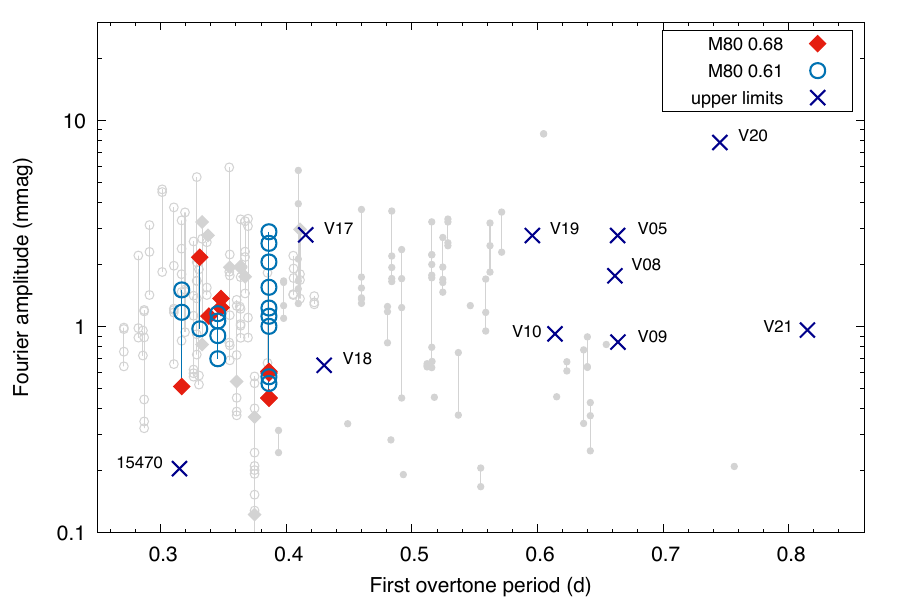}
\caption{Comparison of the Fourier amplitudes of additional-mode signals, against the TESS detections of \citet{Molnar-2021}. The plot clearly shows that except for V17, the K2 data is precise enough to detect sub-mmag signals in the RRc light curves. In contrast, the detection limits are higher for RRab stars, at at least 1--3 mmags. }
\label{fig:mode_amp}
\end{figure}

We then plotted the mode amplitudes and detection limits against the mode amplitudes measured in the TESS sample of \citet{Molnar-2021}. While the difference between the TESS and \textit{Kepler} passbands could introduce some small systematic amplitude differences, in this case we are interested in order-of-magnitude comparisons. As Fig.~\ref{fig:mode_amp} illustrates, the modes we found fit nicely into the spread of RRc mode amplitudes. In V17 and M80\,15470 the detection limits are well below 1.0~mmag, i.e., below most of the signals detected in stars, suggesting that these stars likely do not feature $f_X$ or $f_{0.68}$-type extra modes. On the other hand, the last RRc, V17, could easily contain an undetected low-amplitude mode below 2.5~mmag. 

As for the RRab stars, the detection limits are quite high, between 1--10 mmag for most RRab members. The TESS sample produced only a few modes with relatively low amplitudes in the long-period end of the distribution. We note that extra modes with higher amplitudes have been found in some long-period RRab stars, but those are rare \citep{smolec-2016-rrl,Molnar-2021}. Hence,  the lack of extra modes in the M80 RRab sample may be caused by both the lower sensitivity of the data and by the intrinsic scarcity of extra modes in that period range.

\subsection{Photometric results for other variables}
M80 includes a multitude of different variables beyond the RR Lyrae stars, including bright red giants and various faint stars.

\subsubsection{V01, the type II Cepheid}
\begin{figure}
\includegraphics[width=1.0\columnwidth]{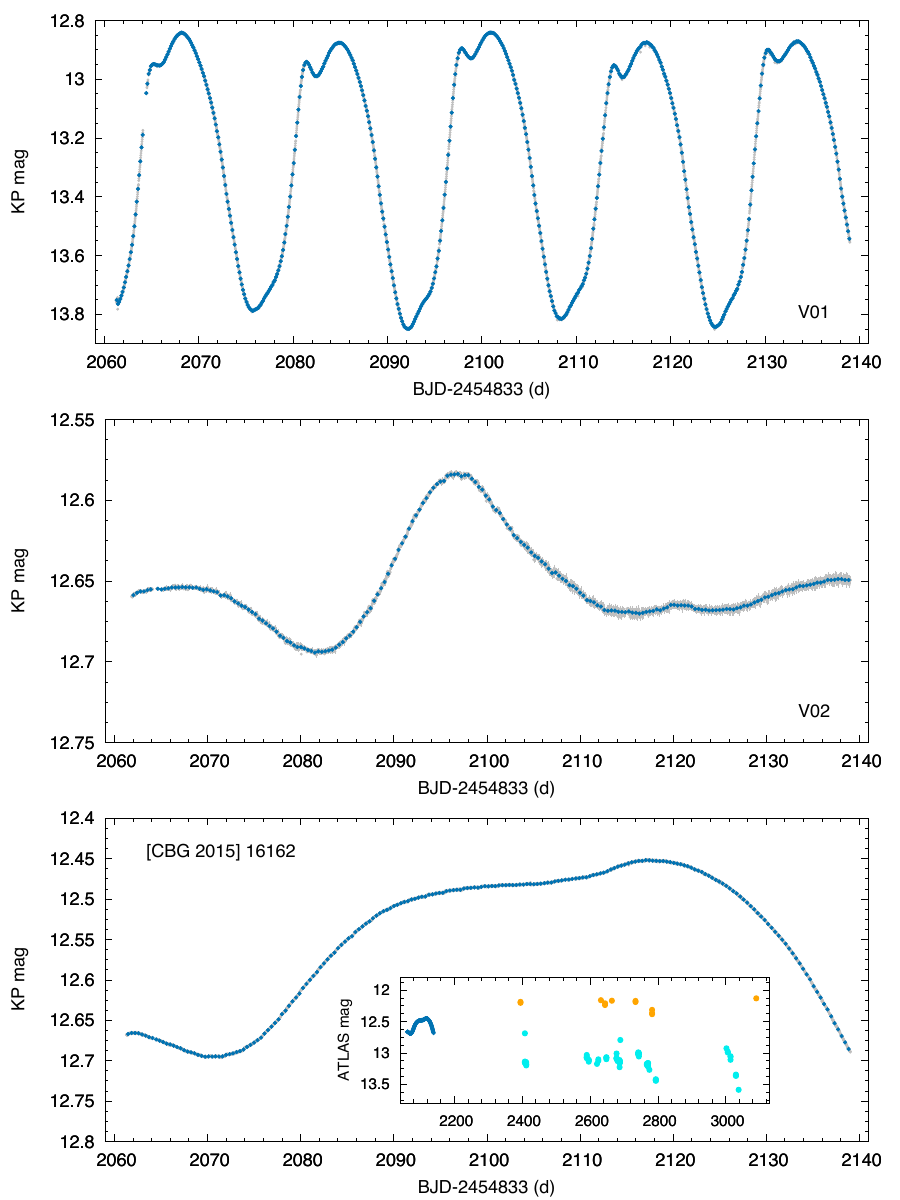}
\caption{Light curves of the three AGB/RGB variables, from top to bottom: V01 is the W~Vir variable, V02 is the known low-amplitude RGB variable, and [CBG~2015]~16162 is the newly identified one. For this last one, the insert shows the K2 light curve together with the ATLAS \textit{c} and \textit{o} (cyan and orange) data. }
\label{fig:rgb-lc}
\end{figure}

This star was analyzed in detail by \citet{Plachy-2017}. Here, we publish an updated version of the light curve that includes the K2SC processing step. We detect the same cycle-to-cycle changes and the corresponding half-integer components. We do not detect any high-frequency components in the star above 2.0~d$^{-1}$, down to 0.25 mmag Fourier amplitudes. The light curve is shown in the upper panel of Fig.~\ref{fig:rgb-lc}. 

\subsubsection{Red giants}
There are multiple long-period variable stars in M80, but most of them have been identified in the core of the cluster by \citet{Figuera-2016}. We were not able to obtain useful light curves for these stars. 

The first known variable red giant in the cluster has been V02, for which a tentative 24.9~d period was suggested by \citet{sawyerhogg-1973}, but was not confirmed later. We detect clear variations in V02, and the first half of the campaign is compatible with a 25~d timescale. The second half of the campaign is, however, ambiguous, and the intrinsic and instrumental variations cannot be separated. The light curve we show in the middle panel of Fig.~\ref{fig:rgb-lc} includes a small correction that we deemed acceptable and mostly affects the last third of the data, raising it slightly in brightness. 

We identified the third variable red giant based on its slow variability in the differential images. The light curve of [CBG~2015]~16162 shows clearer and larger variation than that of V02, but the K2 data is too short for us to determine periodicity. We checked the literature for other observations and found some two-color photometry available from the Asteroid Terrestrial-impact Last Alert System \citep[ATLAS,][]{atlas-2018}. That light curve also shows slow variations and changes in the variation amplitude, but their data is insufficient to determine periodicity, too. In Fig.~\ref{fig:rgb-lc} we show how the K2 data relates to the ATLAS \textit{cyan} and \textit{orange}-filter observations. 

\subsubsection{SX Phe stars}
SX~Phoenicis stars are the Population II counterparts of $\delta$~Scuti-type pulsators. They are blue straggler stars that are formed through binary interaction and are currently crossing the classical instability strip. There are four known SX~Phe stars in the cluster, V22, V23, V24 and V33 plus another candidate, V25, which, according to \citet{Kopacki-2013}, could also be an eclipsing binary. Given their intrinsic faintness and the dense nature of M80, we expected that getting good photometry for them would be difficult. 

\begin{figure}
\includegraphics[width=1.0\columnwidth]{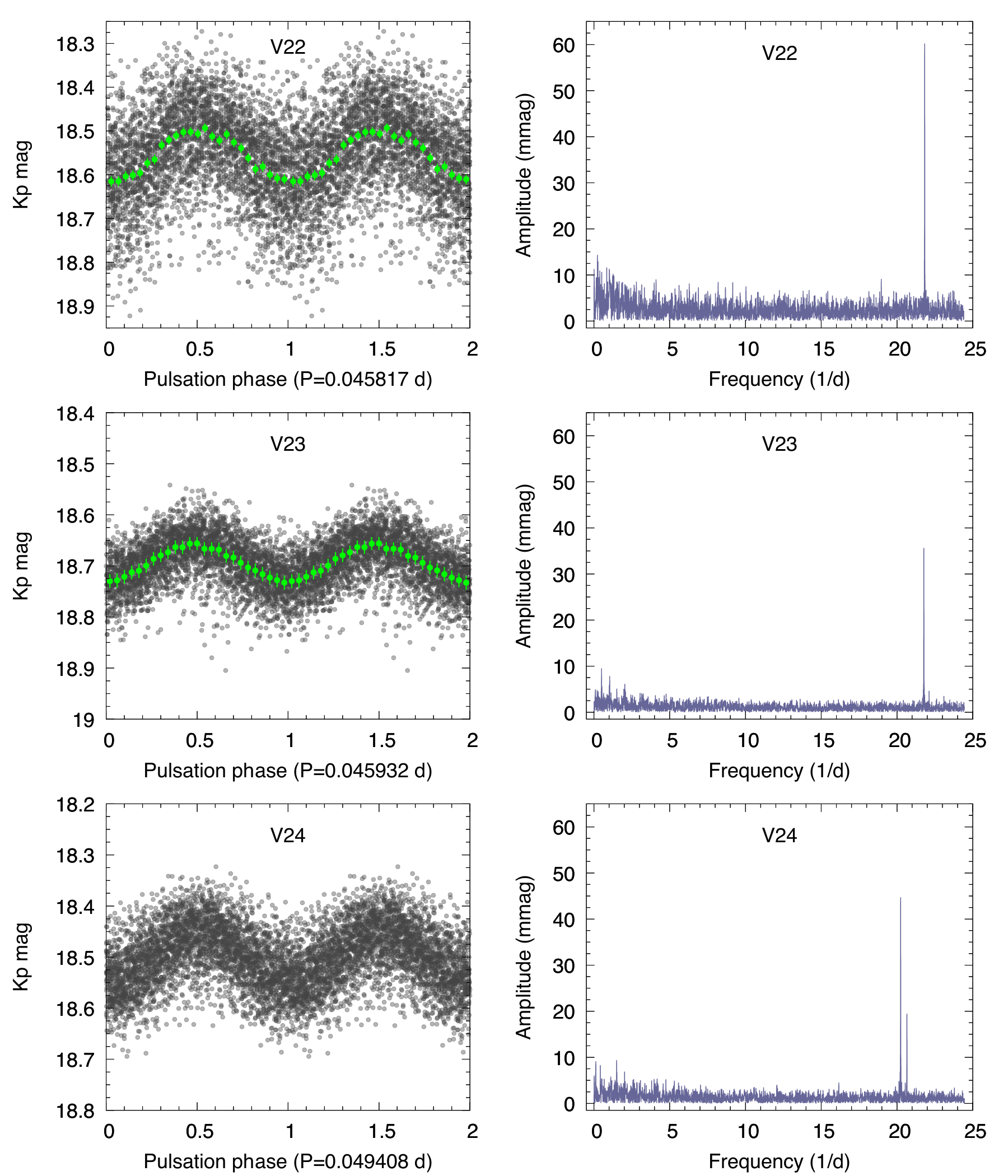}
\caption{Phase curves and Fourier spectra of the disentangled light curves of the confirmed SX~Phe stars. For the monoperiodic stars V22 and V23, binned phase curves are also presented in green. }
\label{fig:sxp}
\end{figure}

V33 was discovered by \citet{Thompson-2010} in the core of the cluster, and we were not able to detect its variation, nor were we able to identify a corresponding \textit{Gaia} source. 

The light curves of V22, V23 and V24 were all strongly contaminated by nearby stars. However, as these pulsate much faster, we were able to separate their variability and generate clean, disentangled light curves for them. In each star, we clearly detected the pulsation components described previously, but we did not detect any new signals. V22 and V23 both exhibit a single pulsation mode, whereas V24 features two modes in close proximity. Their light curves and frequency spectra are shown in Fig.~\ref{fig:sxp}. 

\subsubsection{Binaries and other variables}
\label{sect:eb_ehb}
\begin{figure}
\includegraphics[width=1.0\columnwidth]{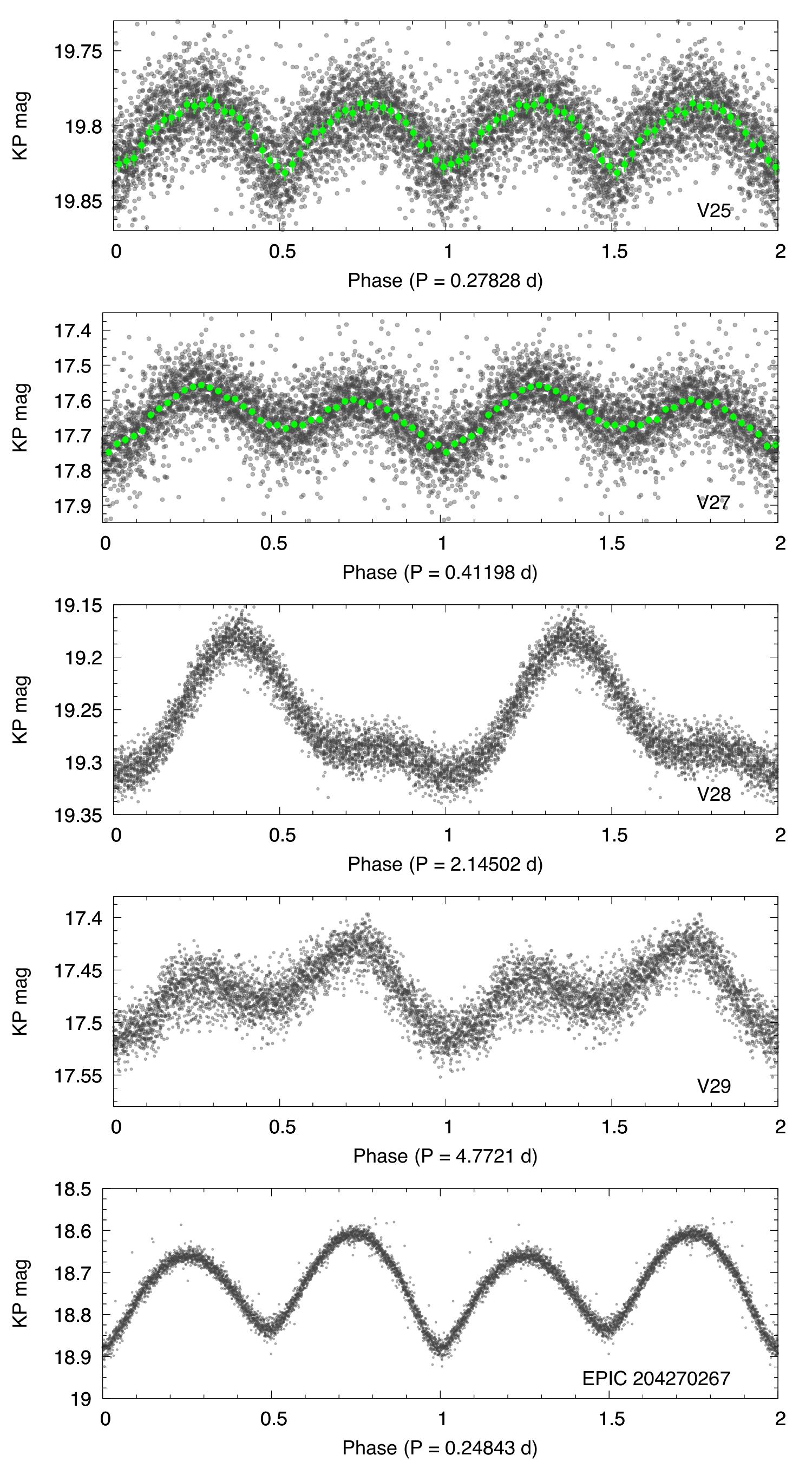}
\caption{Phase curves of the binaries and other variables, including the $\alpha^2$~CVn-type eHB star V28. }
\label{fig:other}
\end{figure}

We extracted light curves from four further known variables (V25, V27, V28 and V29) plus for one new discovery, EPIC~204270267. The folded light curves are presented in Fig.~\ref{fig:other}. 

V25 was originally considered as an SX Phe candidate, and it was the least affected by contamination among those. However, this star turned out to be an eclipsing binary, as suggested by \citet{Kopacki-2013}, with an orbital period of $P_{\rm orb} = 0.278282(3)$~d or 6.6788(7)~hr.

V27 is another eclipsing binary identified by \citet{Kopacki-2013}. We confirm this detection. We also identify another eclipsing binary, EPIC~204270267, which appears to be a foreground system, based on its position in the CMD, as we show in Section~\ref{sec:cmd}. V29 was not classified by \citet{Kopacki-2013}, although based on the brightness and color values they list, it appears to be another foreground object. 

V28 is a more intriguing case. Observations collected by \citet{Kopacki-2013} were insufficient to classify this star, and no color information was recorded either. Based on its position in the \textit{Gaia} CMD (see Sect.~\ref{sec:cmd}), the star appears to be an extreme Horizontal Branch (EHB) star. It was targeted by the spectroscopic study of \citet{moni-bidin_2009}, who derived physical parameters of $T_{\rm eff} = 27600 \pm 700$~K and $\log\,g = 5.38\pm 0.08$ for it, confirming its classification. These stars lost most of their envelopes, and they never evolve to the AGB, but become subdarf B and O stars, before eventually turning into white dwarfs. They can sustain non-radial pusations, but at periods faster than the long cadence sampling of Kepler \citep{brown-2013,randall-2016}. 

Therefore the light curve we observe in V28 must come from some other mechanism: we propose that is caused by $\alpha^2$~CVn-type variation and the 2.1405~d cycle we observe corresponds to the rotation of the star. The changes in surface brightness are caused by chemically peculiar spots on the surface of the star that are maintained for long periods of time by the magnetic fields. Similar variations have been recently observed in a number of EHB stars in three globular clusters, in the 2--50~d period range \citep{momany-2020}. 

\section{Spectroscopic results for RRc stars}
\label{sect:spectro}
The spectra we analyzed had either low resolution or low SNR. Furthermore, since the targets are hot and very metal-poor, very few lines were available for fitting, the H lines dominate the spectra. Here we only fitted the H lines and the continuum, therefore, we were only able to estimate a few parameters: the effective temperatures ($T_{\rm eff}$), the surface gravity ($\log\,g$) and [Fe/H]. Even among those, $\log\,g$ was poorly constrained and was overestimated in some cases. However, this approach was sufficient to estimate the [Fe/H] indices as a change in metallicity causes detectable changes in the continuum normalized flux.

\begin{figure*}
\includegraphics[width=1.0\textwidth]{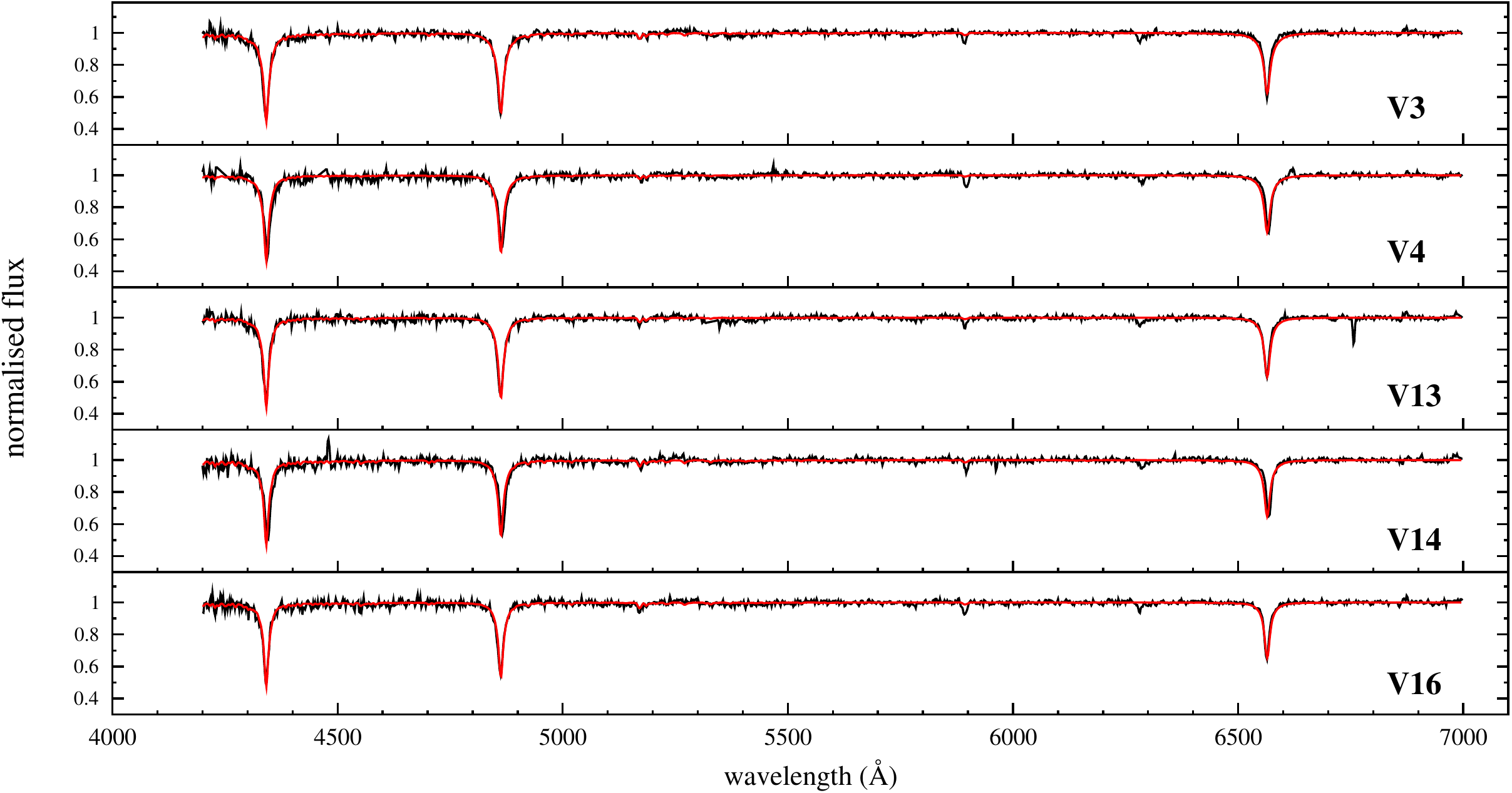}
\caption{The observed Magellan LDSS3 low-resolution spectra (black lines) and best matching synthetic spectra (red lines) of RR Lyrae variables in M80.}
\label{fig:junaspectra}
\end{figure*}

\subsection{The Magellan LDSS3 spectra}

In order to determine the basic physical parameters, we matched the observed spectra with the synthetic spectra using the FERRE code \citep{allende06}. For the grid of synthetic spectra we used the BOSZ models \citep{bohlin17} and convolved new grids that matched the same resolution as the observed spectra covering the wavelength range 4200–7000~\AA. The custom-made grid of model spectra were covering --2.5 < [Fe/H] < --1.0 in steps of 0.25 dex, 3.0 < $\log{g}$ < 5.0 in steps of 0.5 dex, and 6000~K < T$_{eff}$ < 10000~K, in steps of 250 K.

In our analysis we fixed the microturbulent velocity and the value of $\alpha$-elements to 2 km\,s$^{-1}$ and 0.25, respectively, which are reasonable for stars of this spectral type and the low resolution does not enable us to determine these parameters properly from the fit. We shifted the observed spectra to rest wavelengths and used the FERRE code
to quantitatively compare them to each spectrum within the grid of synthetic spectra by minimizing the corresponding $\chi^2$ function. We excluded the following telluric line regions from the fit: 5883--5961 \AA, 6275--6325 \AA, 6475--6523 \AA. 

Fig.\ \ref{fig:junaspectra} shows the observed (black lines) and the best matching synthetic spectra (red lines) of each star. The resulting basic parameters and the estimated random uncertainties are listed in Table\ \ref{table:specparamters}. For determining the uncertainty of the fit, we multiplied the observed spectra by ${\rm 1}+\sigma$ and, ${\rm 1}-\sigma$ which simulates two spectra slightly different from the original spectrum. Then, the fit of each line is repeated while keeping all other parameters unchanged. The average difference between these two new fits and the original best-fit spectrum was the defined uncertainty associated with the fit itself. The low resolution of the spectra and the lack of strong metal lines allows only poor constrains for $\log{g}$ and the metallicity, but the resulting parameters for the latest are in good agreement with the [Fe/H] value of the cluster.

\begin{figure}
\includegraphics[width=0.95\columnwidth]{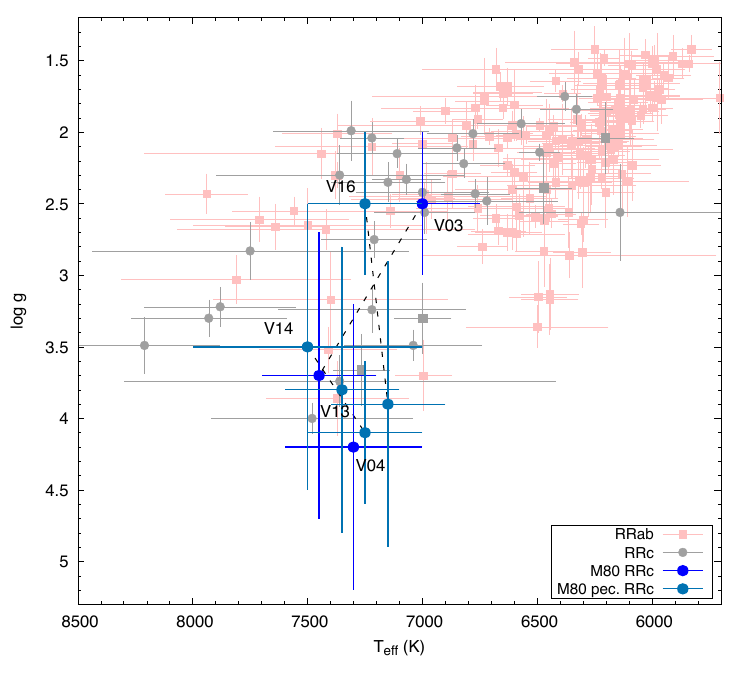}
\caption{Positions of the spectroscopically observed RRc stars in the $T_{\rm eff}$--$\log\,g$ plane, against a large sample of RRab and RRc stars. Dashed lines connect stars where both Magellan and VLT spectra were available.}
\label{fig:teff-logg}
\end{figure}

 \subsection{The archival VLT spectra}
The VLT spectra include a much narrower wavelength range (434--458.9 nm) than the LDSS3 spectra. The stellar atmosphere parameters were estimated via a comparison with a grid of synthetic stellar spectra downloaded from the POLLUX database \citep{Palacios2010}.  The grid included all $\alpha$-enhanced ([$\alpha$/Fe] = $+$0.4) model spectra in the database with 5500 $<$ T$_{eff}$ $<$ 8000 K, 1.0 $<$ log(g) $<$ 4.0, and -3.0 $<$ [Fe/H] $<$ 0.0.  The synthetic models were then smoothed and resampled to match the resolution and binning of the observed spectra.  In preparation for analysis, the observed spectra were iteratively continuum normalized using a low order Chebyshev function and then shifted in wavelength space to match the synthetic spectra via cross correlation.  Intensity residuals in each wavelength bin were recorded for all synthetic and observed spectrum pairs, and then analyzed to determine the best-fit models.

 We were able to get rough parameter estimates for two out of the three observed stars (V03, V14, V16). For V03, we found $T_{\rm eff}$ to be around 6750--7000 K and [Fe/H] of about -2. The former is in agreement with the temperature range of RR~Lyrae stars, while the latter matches the [Fe/H] value of the cluster.

 V16 is also consistent with a low metallicity and appears to be a bit warmer than V03 at 7000--7250 K and with comparable gravity or slightly higher. These differences, however, can also be attributed to differences in pulsation phases during the observations. 

 V14 apprears to have been in a more unstable atmospheric phase of pulsation when the spectrum was taken than the other two stars. This makes the parameters we determined even more uncertain, but the data suggests that it was warmer and had a higher gravity than V03 during observations.

\begin{figure}
\includegraphics[width=1.0\columnwidth]{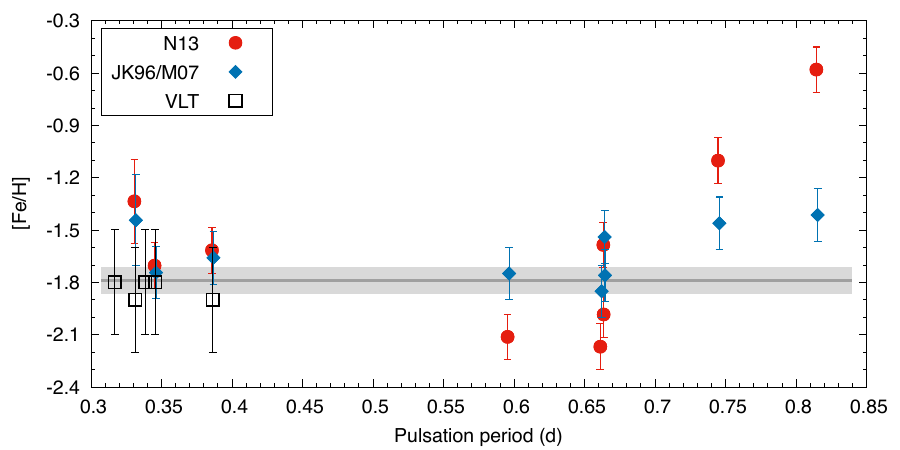}
\caption{[Fe/H] indices of RR Lyrae stars in M80, against their pulsation periods. Black squares show the spectroscopic values from the Magellan observations. Red and blue points show the different photometric estimates. The gray band refers to the spectroscopic [Fe/H] determined from red giants by \citet{Carretta-2015}. Abbreviations N13, JK96 and M07 refer to \citet{nemec-2013}, \citet{JK1996} and \citet{Morgan2007}, respectively.} 
\label{fig:metal}
\end{figure}

We compared the derived $T_{\rm eff}$ and $log\,g$ values to other RR~Lyrae stars, using the parameters published by \citet{nemec-2013} and \citet{crestani-2021} in Fig.~\ref{fig:teff-logg}. The plot shows that although we have high uncertainties, especially in $\log\,g$, our results fit into the distribution of field RRc stars well. The presence of pulsation and the unknown pulsation phases also affect our results, and may explain, at least partially, the differences for stars where we have observations from both telescopes (connected with the dashed lines). Overall, the spectroscopic observations confirm that the parameters of the normal and peculiar RRc stars are similar and that all of them are metal-poor horizontal-branch stars and part of the cluster. 

\subsection{Comparison with photometric [Fe/H] indices}
The $\phi_{31}$ relative amplitude parameter, together with the period, can be used to estimate the photometric [Fe/H] parameter of the stars \citep{JK1996}. One drawback of the relative Fourier parameters---and therefore of the photometric [Fe/H] relations---is that the shape of the light curve is wavelength-dependent. We calculated the photometric [Fe/H] indices for the stars, where we were able to determine the $\phi_{31}$ parameter of the light curve. We used both the relations derived by \citep{nemec-2013} and that were calibrated for the \textit{Kepler} passband, although only for a small sample from the original \textit{Kepler} mission. We also used the original photometric [Fe/H] relations of \citet{JK1996} and \citet{Morgan2007} who first determined them for RRab and RRc stars, respectively, where we corrected for the passband-dependent phase differences following the methods described by \citet{nemec-2013}. Results are listed in Table~\ref{tab:four} in the Appendix. Here we assume an uncertainty of 0.15~dex uniformly, based on the  

Figure~\ref{fig:metal} shows how the various [Fe/H] indices compare to each other. Our spectroscopic values are in good agreement with the values we mentioned in Sect.~\ref{sec:intro}, such as with the results of \citet{Carretta-2015}, which we highlight in the plot with the gray line. For RRc stars, spectroscopic and photometric values agree in most cases except for the peculiar star V13, where the photometric relations overestimate the [Fe/H] index. For RRab stars, however, the spread is quite noticeable. The \citet{JK1996} relation is known to overestimate [Fe/H] for metal-poor stars. The relation of \citet{nemec-2013} can correct for this for short-period stars, but that relation was not calibrated for stars with periods over 0.7~d. As a result, the extrapolated relation starts to deviate steeply towards higher indices for the long-period stars, as Fig.
~\ref{fig:metal} illustrates.

\begin{table} 
\centering 
\caption{Spectroscopic parameters of the observed stars.}
\begin{tabular}{lccc}  
\hline \hline
ID & $T_{\rm eff}$ (K) & $\log{g}$ (dex) & [Fe/H] (dex) \\
\hline
Magellan  & & & \\
V3 & 7450 $\pm$ 250 & 3.7 $\pm$ 1.0 & --1.8 $\pm$ 0.3 \\
V4 & 7300 $\pm$ 300 & 4.2 $\pm$ 1.0 &  --1.9 $\pm$ 0.3 \\ 
V13 & 7350 $\pm$ 250 & 3.8 $\pm$ 1.0 & --1.9 $\pm$ 0.3 \\
V14 & 7250 $\pm$ 250 & 4.1 $\pm$ 0.5 & --1.8 $\pm$ 0.3  \\
V16 & 7150 $\pm$ 250 & 3.9 $\pm$ 1.0 &  --1.8 $\pm$ 0.3 \\
\hline
VLT & & & \\
V3  &   7000 $\pm$ 250    &   2.5 $\pm$ 0.5 &   --2.0 $\pm$ 0.3   \\
V14 &   7500 $\pm$ 500  &   3.5 $\pm$ 1.0   &   --2.0 $\pm$ 0.5  \\
V16 &   7250 $\pm$ 250  &   2.5 $\pm$ 0.5   &   --2.0 $\pm$ 0.3 \\
\hline
\end{tabular}
\label{table:specparamters}
\end{table}

\section{Parallaxes and color-magnitude diagrams}
\subsection{Parallaxes}
\label{sec:parallax}
The \textit{Gaia} mission provides invaluable observations of Milky Way stars and subsystems. We downloaded photometry and parallax data from the Early Data Release 3 (EDR3) for the rectangular area covered by the K2 image mosaic \citep{gaia-edr3-2021}. We then identified the variable stars studied in this paper from the collection of 11230 stars. Since M80 is very dense, not all stars have measurements in the $G_{\rm BP}$ and $G_{\rm RP}$ bands or valid parallax values. Unfortunately this is true for the variables as well: V17, V20, V21, V24 and V27 have no color information. 

We first plotted the parallaxes of the stars to see if any have large values that might indicate a foreground star. We filtered out stars where the RUWE (Renormalised Unit Weight Error) parameter exceeded 1.4, which could indicate either binarity or other issues with fitting a single-star parallax solution. We then corrected for the parallax zero-point offsets with the \texttt{gaiadr3\_zeropoint}\footnote{\url{https://gitlab.com/icc-ub/public/gaiadr3\_zeropoint}} code \citep{Lindegren2021}. 

We were especially interested in the two overluminous RR Lyrae stars V10 and V15, but unfortunately neither of them have useful parallax measurements. V15 have no calculated parallax at all, while for V10 both the parallax uncertainty and the calculated RUWE parameters are exceedingly high. 

\begin{figure}
\includegraphics[width=1.0\columnwidth]{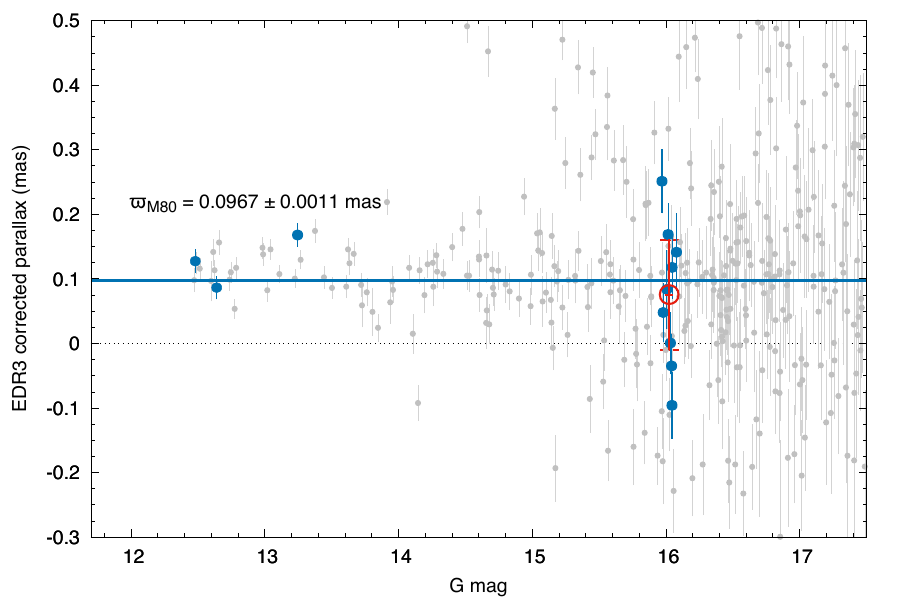}
\caption{Parallaxes of the stars in the field-of-view. Values have been offset-corrected with the method provided by \citet{Lindegren2021}. The blue line marks the parallax of M80. The blue dots are the variable red giants and RR Lyrae stars within the cluster; the red circle is the average parallax of the RR Lyrae group. }
\label{fig:parallax}
\end{figure}

As Fig.~\ref{fig:parallax} illustrates, the parallaxes of the brighter stars, red giants and RR Lyrae stars, agree with the parallax of the cluster itself. Individual parallaxes become rather uncertain at $G > 15$~mag, evident by the spread of the RR Lyrae stars. However, the mean parallax of the nine RR Lyrae stars that have useful measurements is $\overline{\varpi_{RRL}} = 0.075\pm0.085$~mas, which agrees with that of the cluster. 

The rest of the stars are fainter and thus fall into the quickly expanding spread of parallaxes. Of the five further stars for which parallax data are available, two positive and two negative values fall into the general spread of points. Only one star, V22, deviates significantly, with a corrected parallax of $2.51\pm0.31$, which would suggest a distance of only $\sim400$~pc. However, that distance would lead to an absolute brightness of $M_G\approx10.5$~mag, and would put the star well below the main sequence but above the white dwarf region on the Hertzsprung--Russell diagram, which is almost entirely unpopulated. On the other hand, the variation of the star and its position on the color-magnitude diagram (see next section) of the cluster both suggest that it is an SX~Phe star within the cluster. We therefore conclude that it is more likely that its parallax has been measured erroneously.  

\subsection{Color-magnitude diagrams}
\label{sec:cmd}

\begin{figure*}
\includegraphics[width=1.0\textwidth]{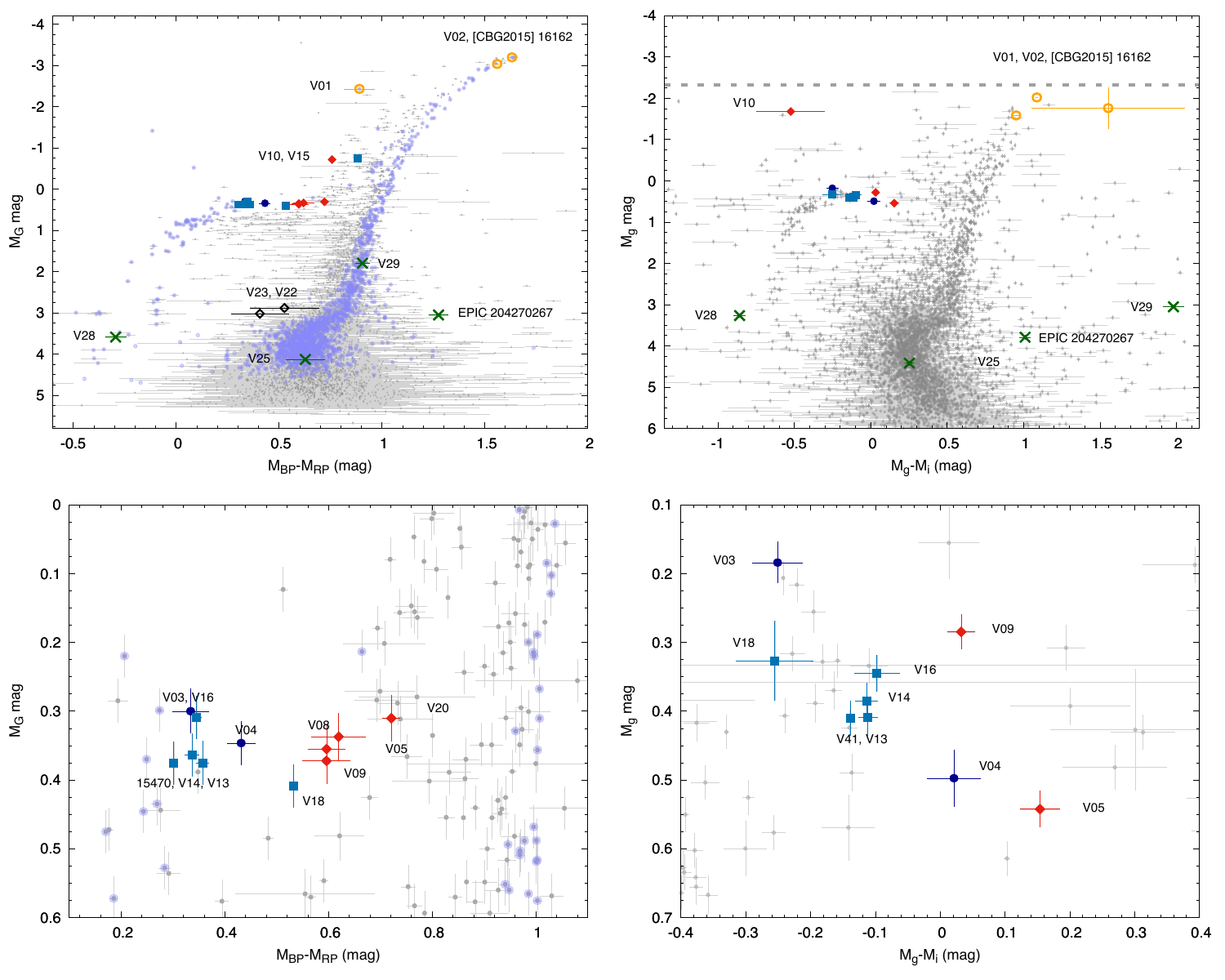}
\caption{Absolute color-magnitude diagrams. Left: based on \textit{Gaia} EDR3. gray points are all the stars in the K2 image mosaic. The cleaned CMD is shown with the blue points. Variables are indicated with larger symbols. Right: CMD based on the Pan-STARRS DR1 survey. The bottom plots show the areas around the RR Lyrae instability strip. Dark and light blue points are the regular and peculiar RRc stars, respectively; RRab stars are shown in red.  }
\label{fig:cmds}
\end{figure*}

With the \textit{Gaia} EDR3 photometry at hand, we constructed a color-magnitude diagram (CMD) for the cluster. First, we simply queried the area of the K2 mosaic image, in order to contain all stars, including variables that might not be part of the cluster. Here we only excluded stars where $G_{BP}$ and/or $G_{RP}$ data were not available or their flux-over-error ratio was below five in those passbands. We then also created a clean CMD to show where the population of the cluster really lies. For this, we first filtered stars by proper motion, removing those that have no data or differ from the proper motion of the cluster by more than 2.2~mas/yr. This limit was chosen based on the radial profile of the cluster in proper motion space. We then filtered the CMD further based on the \textit{Gaia} corrected color excess factor parameter, $C^*$. This factor indicates when there is an imbalance between the three different fluxes measured by \textit{Gaia}, but a high score does not necessarily indicate poor data quality \citep{riello-2021}. A higher $C^*$ cutoff results in higher completeness at the cost of lower purity. Here chose a limit of $\pm 10\sigma_{C^*}$ to obtain a clean but still adequately populated CMD for the cluster, following Eq.~18 of \citet{riello-2021}, to take the increasing photometric scatter towards fainter stars into account.The final filtering step was the exclusion of stars where $G_{BP} \ge 20.3$ mag. This is needed because towards the faint end of the \textit{Gaia} dataset, the $G_{BP}$ fluxes tend to be overestimated, which can cause systematic bias for red sources \citep{riello-2021}. These filters removed most of the faint stars on the Main Sequence, and some of the strongly variable stars as brightness changes also elevate the $C^*$ scores. 

To transform the apparent \textit{Gaia} brightnesses to absolute magnitudes, we used the distance modulus provided by \citet{Baumgardrt-2021} calculated the interstellar extinction with the \texttt{mwdust} code for each \textit{Gaia} passband \citep{bovy2016}. The extinction values are: $A_{\rm G} =  0.599 \pm 0.020$~mag; $A_{\rm BP} = 0.715 \pm 0.025$~mag; $A_{\rm RP} = 0.414 \pm 0.025$~mag. The absolute CMD for the cluster, with the variable stars indicated, can be seen in Fig.~\ref{fig:cmds}.

As comparison, we also downloaded the mean photometry of sources up to 4' distance from the center of the cluster from the second release of the Pan-STARRS $3\pi$ survey \citep{Chambers2016}. The Pan-STARRS photometry is affected by crowding and saturation towards the core: this can be seen in the left-hand image of Fig.~\ref{fig:gc-map}, but also in the distribution of detections in sky coordinates, which includes multiple string-like features which do not correspond to real stars. While such spurious detections can be filtered out, these issues lead to poor coverage of the core. Furthermore, Pan-STARRS photometry saturates around 13--14 mag, which effectively cuts off the brightest red giants from the CMD: we mark the $g$-band saturation limit with the dashed line on the right side of Fig.~\ref{fig:cmds}. The saturation limit clearly affects the variable red giants in the sample as well, with V01 and [CBG2015]~16162 appearing much redder than in the \textit{Gaia} CMD. For this CMD, we used the averages from the \citet{SFD-1998} and \citet{S-F-2011} dust and reddening maps\footnote{NASA/IPAC Infrared Science Archive, \url{https://irsa.ipac.caltech.edu/}}: $A_g = 0.75 \pm 0.05$~mag and $A_i = 0.40 \pm 0.04$~mag.

We highlighted the positions of the RR~Lyrae stars in the HB from both CMDs in the lower panels of Fig.~\ref{fig:cmds}. These clearly show that the RRab stars, the normal RRc stars and the strongly modulated RRc stars form a continuous group in the HB. This confirms that the modulated stars are indeed RR Lyrae stars, despite the unexpected light curve shapes. 

In fact, the spread of RR~Lyrae stars is very narrow, only about 0.1~mag in absolute brightness, indicating extremely similar luminosities for the entire RR~Lyrae population. In contrast, the CMD of about a hundred field stars presented by \citet{Molnar-2021} and the CMD of the entire \textit{Gaia} DR3 sample published by \citet{clementini-2022} both span more than 1.0 magnitude in $M_{\rm G}$ brightness. 

With the \textit{Gaia} CMD, we can also confirm the positions and thus evolutionary stages of two out of the three SX~Phe stars. V22 and V23 are indeed blue stragglers above the MS turnoff point. 

The newly identified binary EPIC~204270257 appears to lie redward both from the MS and the RGB. It is very likely that this is a foreground star, which is intrinsically fainter than our CMD would suggest. A look at the single- and binary-star main sequences on the Gaia HRD indicates that a binary with identical components at the same $G_{\rm BP}-G_{\rm RP}$ color would be about three magnitudes fainter at $M_G \sim 6$~mag. That difference would put it to about the quarter of the distance of M80 \citep{gaiadr2-hrd-2018}. 

A notable star in the CMDs is V28, which appears to be an intrinsically very blue but not very luminous object. Assuming that it is a cluster member, then it is a hot subdwarf star along the EHB, the blue, downward extension of the HB, as we mentioned in Sect.~\ref{sect:eb_ehb} \citep{Culpan-2022}. The observed rotational variation rules out the possibility of it being a foreground white dwarf, as those do not have star spots and thus would not produce such light curves \citep{kawaler-2004}. EHB stars, on the other hand, have been recently found to show chemical spots, as we mentioned earlier: therefore we classify V28 as a rotational or $\alpha^2$~CVn-type variable within the cluster.

\subsection{MIST Isochrone fits in multiple photometric systems}
\label{sec:MIST}



We use the publicly available MIST database \citep{MIST1} to generate non-rotating isochrones spanning ages 10 to 17 Gyr. These are transformed to the magnitudes and colors of the relevant photometric systems corresponding to three sources considered: Bessell \textit{UBVRI} \citep{Bessell-1990}, reproducing the Johnson-Cousins system, to accompany photometry from \citet{Alcaino1998}, Pan-STARRS \textit{grizy} passbands for the PS1 data, and Gaia \textit{G}, $G_{\rm BP}$, $G_{\rm RP}$ passbands for EDR3 to accompany data queried in the present analysis.
In determining the best-fitting age and metallicity for M80, we consider isochrones with [Fe/H] values of $-1.2$ to $-2.1$ dex in increments of $0.1$ dex.

We first re-fit the data presented by \citet{Alcaino1998} using the much more recent MIST models. In their analysis, \citet{Alcaino1998} combine inferences of 16 Gyr and 18 Gyr using two different sets of isochrones---those of \citet{Bergbusch1992} and \citet{McClure1987}, respectively---to report a best-fitting composite age and metallicity of $17\pm2$ Gyr and [Fe/H]$=-1.70 \pm 0.15$.


\begin{table*} 
\centering 
\caption{Summary of derived parameters for M80 based on isochrone fitting}
\begin{tabular}{ll c c l }  
\hline\hline
Data Source & Color Transform & Age (Gyr)$^{\star}$ & [Fe/H] (dex) & Comments \\
\hline
\citet{Alcaino1998} & \textit{UBVRI}  & $12-17$ & $-1.7 \pm 1$ & adjustments to the distance modulus and both\\
& & & &  extinction coefficients are required (see Fig \ref{fig:compare_iso_fits}) \\
PS1 DR2             & PanSTARRS & $14 \pm 1(\pm3)$  & $-2.0 \pm 0.1$ & higher metallicities require high ages; ex. [Fe/H]=$-1.8$ \\
& & & &  and 17 Gyr; adjustment not strictly required \\
Gaia DR2            & Gaia DR2  & $13\pm1(\pm3)$  & $-1.8\pm0.1$ & older ages are better fit by lower metallicities; \\
 & & & & adjustment to $A_{BP}$ is required for alignment \\
Gaia EDR3           & Gaia DR2  & $14\pm1(\pm3)$ & $-1.75 \pm 0.15$ & adjustment to extinction coeffs not strictly required \\
 \hline
\end{tabular}
\tablefoot{ $^{\star}$Note that the uncertainties quoted for isochrone-based ages (e.g. 1 Gyr) are a rough reflection of uncertainty due to ambiguity in the best-fitting isochrone, among those presented, by visual fit only. Global uncertainties for isochrone-based age determinations for even the most well-studied and well-constrained globular clusters are likely 2 or 3 Gyr (see, e.g., \citealt{Joyce2023}). This is reflected parenthetically, e.g. $\pm 1 (3)$ Gyr. }
\label{table:parameter_summary}
\end{table*}

\begin{figure*}
     \begin{subfigure}[b]{0.33\textwidth}
         \centering
         \includegraphics[width=\textwidth]{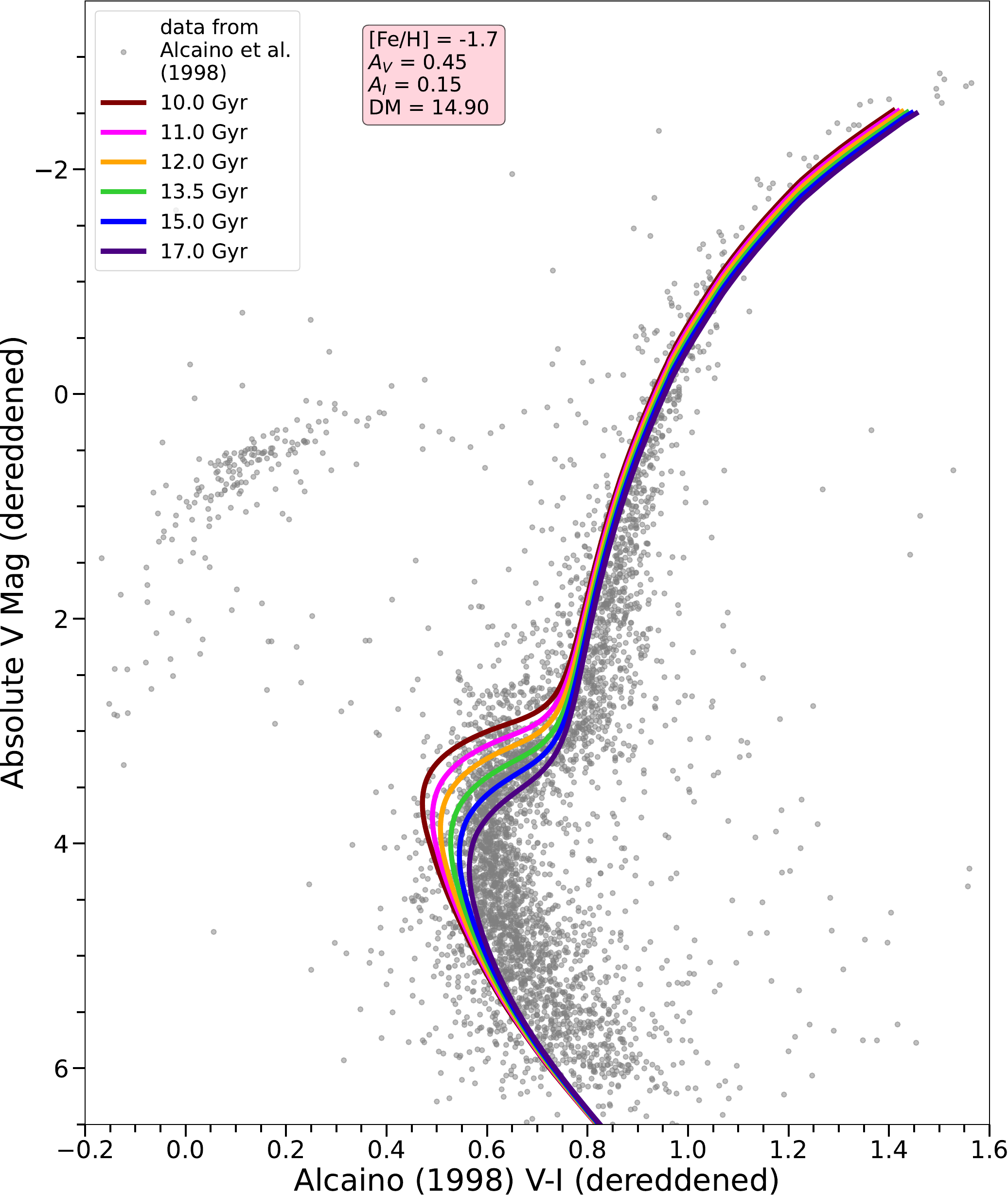}
         \caption{\citet{Alcaino1998} }
     \end{subfigure}
     \hfill
     \centering
     \begin{subfigure}[b]{0.33\textwidth}
         \centering
         \includegraphics[width=\textwidth]{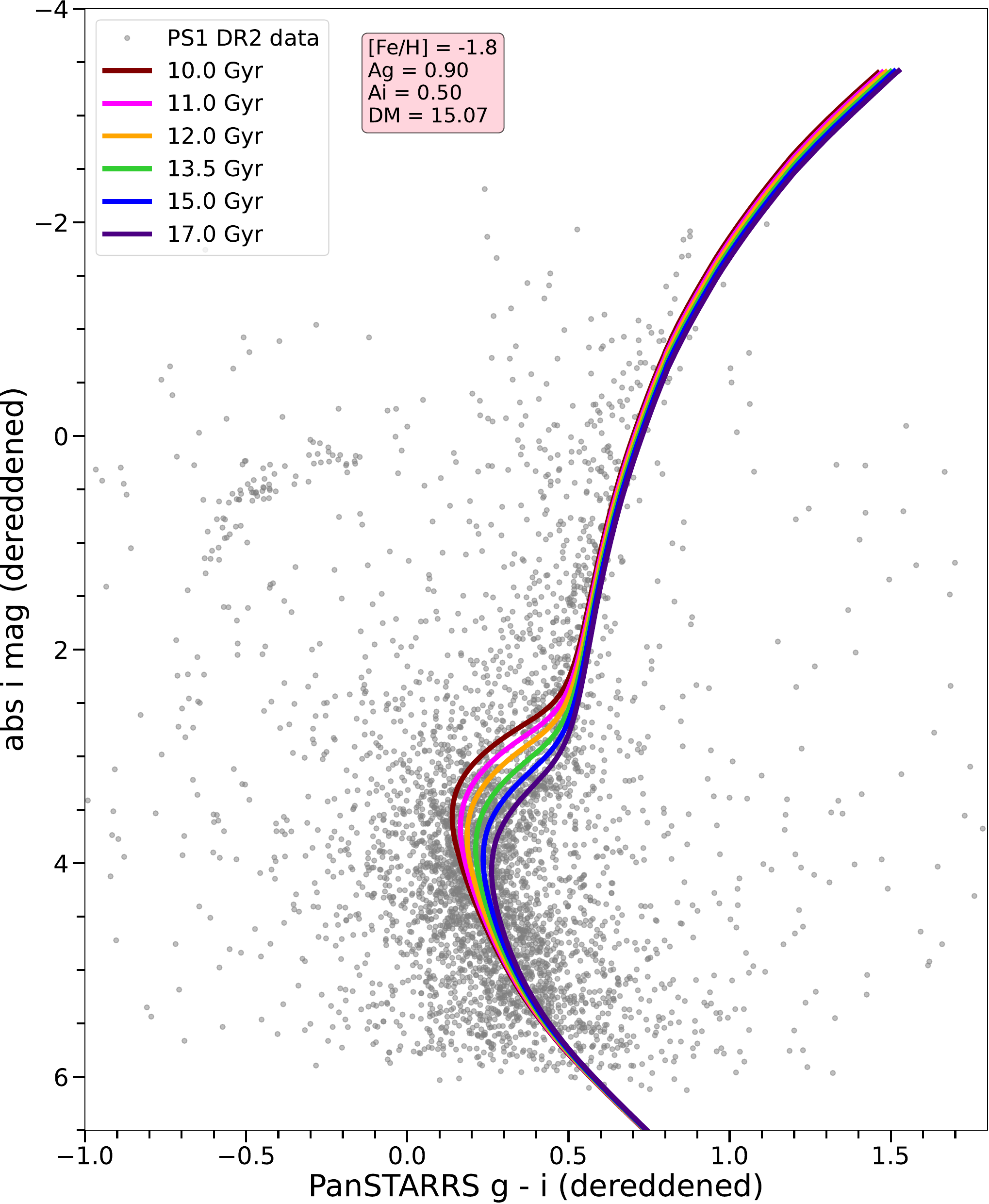}
         \caption{PanSTARRS}
     \end{subfigure}
     \hfill
     \begin{subfigure}[b]{0.33\textwidth}
         \centering
         \includegraphics[width=\textwidth]{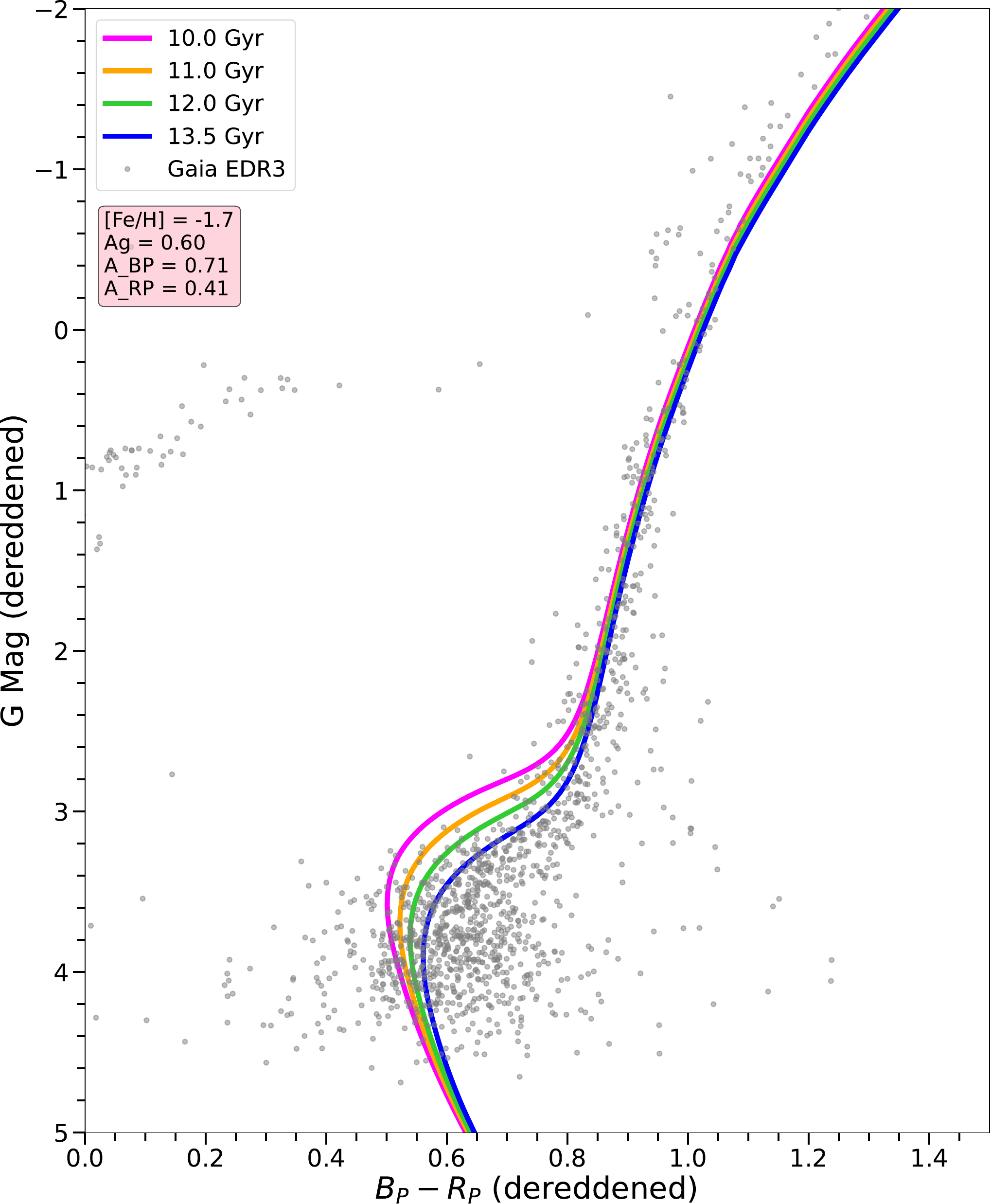}
         \caption{\textit{Gaia} EDR3}
     \end{subfigure}
        \caption{MIST isochrone fits to each data source. Horizontal bars are as described in \ref{sec:MIST}. }
        \label{fig:compare_iso_fits}
\end{figure*}

Presently, we find that MIST isochrones with a metallicity of [Fe/H]$=-1.7 \pm +0.1$, under the assumption of no $\alpha$--element enhancement ([$\alpha$/Fe]$=0$), still fit \citet{Alcaino1998}'s data well, particularly along the red giant branch. However, models with ages in the range of 12 to 13 Gyr show a level of consistency similar to that of 17 Gyr models. We also find that for any combination of age and metallicity considered, some adjustment to the distance modulus and reddening is required to achieve alignment between the data and models---this cannot be accomplished by varying the isochrone input parameters alone.
The first panel of Figure \ref{fig:compare_iso_fits}shows a set of MIST isochrones with [Fe/H]$=-1.7$ and ages as indicated overlaid on the data of \citet{Alcaino1998}. To provide the best visual fit, V-band and I-band extinction values of  $A_V = 0.568$ and $A_I = 0.334$, respectively, and a distance modulus of $(m-M)=14.96$ are adopted.

Though there is some tension between our results and those of \citet{Alcaino1998}, we note that both sets of models used in their analysis are 30 (or more) years old.
In addition to this, the isochrones of \citet{Bergbusch1992} were oxygen-enhanced, but \citet{Alcaino1998}'s results were derived prior to the revision of solar abundances in 2009 \citep{A09}. Likewise, the helium contents adopted in the models of \citet{McClure1987} were known to be inappropriate by the time of \citet{Bergbusch1992}'s publication. 
That estimations of the fundamental parameters of stellar populations can shift so much on the basis of abundances underscores the importance of treating these quantities carefully in stellar modeling.

More recent literature fits of isochrones to M80 include those of \citet{BarkerPaust2018}, who report best-fitting metallicities and ages of M80 according to three different sets of isochrones:  
\begin{itemize}
\item[--] [Fe/H]$=-1.65$ dex, $12.5$ Gyr using DSEP (the Dartmouth Stellar Evolution Program, \citealt{Dotter2008});
\item[--] [Fe/H]$=-1.68$ dex, $13.5$ Gyr using PARSEC (the PAdova and TRiest Stellar Evolution Code, \citealt{PARSEC}); and
\item[--] [Fe/H]$=-1.90$ dex. $14.0$ Gyr using MIST.
\end{itemize}

We now fit MIST isochrones, computed over the same parameter ranges, to two new data sources. 
The center and right-hand panels of Figure \ref{fig:compare_iso_fits} show the best-fitting models to M80 data from (1) PanSTARRS DR2 mean photometry and
(2) \textit{Gaia} EDR3, respectively.
Table \ref{table:parameter_summary} summarizes the age and metallicity determinations for each data set. We note that in three cases (including \citealt{Alcaino1998}), the model-derived metallicities are reported under the assumption that [$\alpha$/Fe]$=0$. We note that these can be rescaled to an $\alpha$-enhanced basis (e.g., [$\alpha$/Fe]$=+0.3$) as needed. For example, the combination [Fe/H]$=-1.8$, [$\alpha$/Fe]$=0.3$ is roughly equivalent to [Fe/H]$=-1.59$, [$\alpha$/Fe]$=0$, and so forth (see Table 1 of \citealt{Joyce2023} and related discussion on how to account for $\alpha$-enhancement in MIST).  
We find that the age and metallicity determinations for M80 based on modern data are consistent with the MIST results presented in \citet{BarkerPaust2018} and broadly consistent with each other.

The fact that there is some variance among best-fitting ages and metallicities for the same targets, using the same models, perhaps calls into question consistency between the extinction maps used to distance-correct the three different observational data sets. However, this also demonstrates that astronomers should proceed with caution when using isochrones--which are themselves interpolated models subject to their own complex network of uncertainties--to perform age determinations. The global precision with which isochrone-based ages can be stated, realistically, is at the 2-3 Gyr level (including considerations for modeling uncertainties and variations in physical assumptions; see \citealt{Joyce2023, JoyceTayar2023}). By this metric, our age determinations according to \textit{Gaia} and PanSTARRS are formally consistent, and the age determinations according to \citet{Alcaino1998} are approximately $1~\sigma$ older.

\section{Conclusions}
In this paper, we analyzed the K2 observations of the globular cluster M80. Despite the high stellar density, we were able to recover the variations of most known variables in the cluster, and discovered two new ones. We successfully detected modulation and low-amplitude extra modes in RR Lyrae stars. The distribution of RRc extra modes in clusters, i.e., in a population with a narrow metallicity spread, make it possible to compare them against pulsation models more accurately, with the metallicity being constrained. \citet{Molnar-2021} and \citet{netzel-2023} found that pure RRc stars are prevalent among stars bluer than 0.30--0.33 mag in dereddened BP--RP color, based on TESS and K2 observations, respectively. Although the detection limits in M80 varied from star to star, we find that our detections are consistent with that of the field stars: out of the three stars where we did not find f$_{0.61}$ modes, two are at the blue edge, at around BP--RP\,$\approx$\,0.3.

\subsection{A new RR Lyrae subclass?}
\label{sect:newclass}
Beyond the extra modes, the light curve shapes of RRc stars present an interesting dichotomy as well, with some stars (especially V03 and V04) appearing as normal overtone variables with clear hump features, whereas others showing very sinusoidal variations with strong modulation(s) (V13, V14, V16 and M80~15470). Our photometric and spectroscopic results confirm that these stars are members of the cluster and hence are in the same position as the regular RRc stars. 

The peculiar RRc stars are characterized by (at least) two close frequencies with similar amplitudes that are within an order of magnitude to each other, and a light curve that lacks the characteristic hump features just before maximum light. They might be a subtype of BL1 stars, i.e., RRc stars where we observe only one side-peak next to the pulsation frequency instead of modulation triplets. Examples of such stars exist elsewhere, too, but we know little else about them. The most prominent examples are V701~Pup \citep{antipin-2005IBVS}, the two Bulge stars identified by \citet{netzel2019}, and V37 in the globular cluster NGC6362 \citep{smolec-2017}. We searched for more such stars in the literature, but light curves are often sampled too sparsely for clear classification. We identified the variable star KT--26 in M22 \citep{Rozyczka-2017}, V12 in NGC6401 \citep{Tsapras-2017}, V8, V13 and V16 in NGC4147 \citep{arellano-2018} as further examples of peculiar, strongly phase- and amplitude-modulated RRc stars with sine-like light curves.

One reason for the relative obscurity of these stars might be that classification pipelines easily miss these or assign them to different classes (e.g., rotational variables) since their light curves are unlike regular RR Lyrae stars. As such, it is unclear how frequent they might be among RR Lyrae stars, and if they warrant a separate subcategory or not. This could be answered by combining the capabilities of the TESS and \textit{Gaia} missions, where absolute CMD positions can help to find more RR Lyrae stars with unusual light curves.   

Our results highlight the rich landscape of RR Lyrae pulsations, especially that of RRc stars. These stars feature not only an abundance of extra modes, but various types of modulations as well: the classical Blazhko-type in BL2 stars with symmetric side-peaks in the frequency spectra; the beating-like variation in BL1 stars between distinct, close-by frequencies; and the third group of phase-modulated RRc stars (not represented in M80) that cannot be explained with simple light-time effects \citep[see, e.g.,][]{derekas-2004,sodor-2017}.

\subsection{New and newly classified variables}
We identified one new RRc star and one new long-period red giant variable star in the cluster. The RRc star is a new member of the low-amplitude, strongly modulated group present in the cluster. 

Regarding the other variables, the K2 campaign unfortunately was too short to determine periodicities in red giant variables. On the other hand, we confirm that the variables V25 and V27 are eclipsing binaries. Finally, V28 turned out to be an interesting case: its position in the cluster CMD and its variation indicates that it is an EHB star exhibiting rotational variation. The star likely features chemically peculiar spots on its surface, similar to other EHB variables \citep{momany-2020}.

\subsection{The case for globular cluster seismology}
Few asteroseismic studies focused on globular clusters so far. This is understandable given their large distances and compact nature, and the corresponding resolution and sensitivity limitations of space missions like \textit{Kepler} and TESS. However, as both our study and the works of \citet{wallace-m4variables-2019,wallace-ularrl-2019}, \citet{Howell-2022,howell-2023}, or \citet{tailo-2022} show, continuous and high-precision photometry can reveal a wealth of information about the clusters and their constituent stars. The analysis of clusters was identified as one of the important topics where archival K2 data can deliver new discoveries \citep{barentsen-2018}. Analysis of further clusters observed by K2 is under way (e.g., Kalup et al., in prep., Howell et al., in prep.)

Globular clusters are not well-suited targets for photometry missions like TESS and PLATO that work with small telescopes that have large fields of view at the cost of low angular resolution (21 and 15 "/px, respectively). Therefore, any observations of even the closest and loosest clusters like M4 will provide only limited scientific results. There are, however, larger missions in development as well. 

The European HAYDN (High-precision AsteroseismologY in DeNse stellar fields) project is an ESA Voyage 2050 mission proposal, and the only one that is being designed specifically for asteroseismic purposes \citep{miglio-haydn-2019}. The prospective space telescope is designed to survey dense stellar fields like the bulge, as well as open and globular clusters. It is planned to have an optical system similar to \textit{Kepler} in size, but with a much higher angular resolution (0.25"/px). This mission is expected to detect oscillations down to the base of the red giant branch for multiple clusters. 

Another proposed mission, the Chinese \textit{Earth 2.0} (or ET) project that is planned to have one 30\,cm telescope pointed towards Baade's window in the Bulge in order to search for microlensing events \citep{et-mission-2022}. This microlensing telescope will have sub-arcsecond pixel resolution but only over a 4\,deg$^2$ field of view. Hopefully this will still include the two globular clusters within Baade's window, NGC~6522 and NGC~6528. 

The largest telescope that is planned to aim towards the Bulge repeatedly is the American \textit{Nancy Grace Roman} Space Telescope, funded by NASA. A significant amount of telescope time will be dedicated to the Galactic Bulge Time Domain Survey, which will be carried out in a way that is very similar to the K2 campaigns and thus offers many comparisons \citep{Roman-2019,Roman-bulge-2022}. \textit{Roman} will observe the selected 2\,deg$^2$ Bulge fields continuously for 62 days in each run with 15\,min cadence, but at a considerably better resolution and depth (0.1\,"/px and down to 22--24 mag, respectively) in red and near-infrared wavelengths. Although the survey is currently optimized for microlensing, the inclusion of any globular cluster within or near the Bulge would be highly desirable, as they can provide a huge variety of stellar and galactic science cases \citep[see the white papers of][]{molnar_wp-2023,grunblatt-2023}. 

These future capabilities would allow us to study more clusters and in much more detail than what we were able to achieve with \textit{Kepler}.

\begin{acknowledgements}
       This research was supported by the `SeismoLab' KKP-137523 \'Elvonal grant of the Hungarian Research, Development and Innovation Office (NKFIH). The research leading to these results has received funding from the LP2012-31, LP2018-7 and LP2021-9 Lend\"ulet grants of the Hungarian Academy of Sciences. Cs.~Kalup was supported by the \'UNKP-21-2 and \'UNKP-22-2, H.~Netzel by the \'UNKP-22-4 New National Excellence Programs of the Ministry of Innovation and Technology from the source of the National Research, Development and Innovation Fund. M.J. gratefully acknowledges funding of MATISSE: \textit{Measuring Ages Through Isochrones, Seismology, and Stellar Evolution}, awarded through the European Commission's Widening Fellowship. This project has received funding from the European Union's Horizon 2020 research and innovation program. This work was supported by the PRODEX Experiment Agreement No. 4000137122 between the ELTE E\"otv\"os Lor\'and University and the European Space Agency (ESA-D/SCI-LE-2021-0025).

       This paper includes data collected by the K2 mission. Funding for the \textit{Kepler} and K2 missions is provided by the NASA Science Mission Directorate. This work has made use of data from the European Space Agency (ESA) mission {\it Gaia}, processed by the {\it Gaia} Data Processing and Analysis Consortium (DPAC). Funding for the DPAC has been provided by national institutions, in particular the institutions participating in the {\it Gaia} Multilateral Agreement. Based on data obtained from the ESO Science Archive Facility with DOI:
\url{https://doi.eso.org/10.18727/archive/27}. This research made use of NASA’s Astrophysics Data System Bibliographic    Services, and of the SIMBAD database and "Aladin sky atlas", operated at CDS, Strasbourg, France.

       LM thanks the hospitality of the Space Telescope Science Institute (Baltimore, MA), where part of this project was carried out. Fruitful discussions with Dr.\ Rachael Beaton and Dr.\ Annalisa Calamida are gratefully acknowledged.

\end{acknowledgements}

%
%

\bibliographystyle{aa}
\bibliography{m80-k2}{}

\begin{appendix}
\section{Data tables}
\label{app}

\begin{table*}
\centering 
\caption{Corrected K2 photometry of the variable stars in M80. The entire table is available as a machine-readable table online.}
\label{tab:lc}
\begin{tabular}{lccccc}  
\hline \hline
ID & \textit{Kepler} BJD & \textit{Kp} (mag) & $\sigma_{Kp}$ (mag) & $F_{\rm corr}$ (e/s) & $\sigma_F$ (e/s) \\
\hline
V01 & 2061.264820 & 13.754550 & 0.003100 & 41512.757 &  118.573\\
V01 & 2061.285252 & 13.760330 & 0.001750 & 41292.350 &  66.541\\
V01 & 2061.305684 & 13.750510 & 0.001780 & 41667.321 &  68.396\\
V01 & 2061.326116 & 13.754120 & 0.001770 & 41529.019 &  67.665\\
V01 & 2061.346548 & 13.751600 & 0.001770 & 41625.573 &  67.761\\
V01 & 2061.366980 & 13.747290 & 0.001780 & 41791.279 &  68.473\\
V01 & 2061.387412 & 13.746130 & 0.001780 & 41835.706 &  68.383\\
V01 & 2061.407845 & 13.743240 & 0.001780 & 41947.125 &  68.604\\
\multicolumn{6}{l}{\dots}\\
\hline
\end{tabular}
\end{table*}

The appendix includes various data tables. The five wavelength-calibrated LDSS3 spectra are available as supplementary FITS tables. Table~\ref{tab:lc} presents the K2 light curves. Tables~\ref{tab:amp} and \ref{tab:phase} list the periods and the Fourier amplitudes and phases of the pulsation frequency and its first four harmonics for the RR Lyrae stars. The epoch for the phases is the \textit{Kepler} epoch of BJD\,2454833.0. Table~\ref{tab:four} lists the relative Fourier parameters and the estimated photometric [Fe/H] indices of the RR~Lyrae stars in the cluster. Table~\ref{tab:absmag} lists the \textit{G}-band absolute magnitudes, dereddened BP--RP colors and zeropoint-corrected parallaxes based on the \textit{Gaia} EDR3 catalog \citep{gaia-edr3-2021}. 

\begin{table*}
\centering 
\caption{Fourier amplitudes of the M80 RR Lyrae stars.}
\label{tab:amp}
\begin{tabular}{lcccccc}  
\hline\hline
ID & P   & $A_1$ & $A_2$ & $A_3$ & $A_4$ & $A_5$  \\
 ~ & (d) & (mag) & (mag) & (mag) & (mag) & (mag)   \\
\hline
\multicolumn{7}{c}{RRab stars}\\
V05 & 0.663933 & 0.2262(6) & 0.1162(6) & 0.0727(6) & 0.0430(6) & 0.0196(6) \\
V08 & 0.661808 & 0.2182(5) & 0.1185(4) & 0.0736(5) & 0.0498(5) & 0.0258(5) \\
V09 & 0.664178 & 0.2214(4) & 0.1183(4) & 0.0755(4) & 0.0483(4) & 0.0251(4) \\
V19 & 0.596075 & 0.0852(6) & 0.0410(5) & 0.0237(6) & 0.0133(6) & 0.0072(5) \\
V20 & 0.745276 & 0.14806(12) & 0.06205(13) & 0.03802(12) & 0.01479(10) & 0.00697(14) \\
V21 & 0.815007 & 0.0709(2) & 0.0293(2) & 0.0136(2) & 0.0047(2) & 0.0029(2) \\
\hline
\multicolumn{7}{c}{RRc stars}\\  
V03 & 0.345711 & 0.16795(12) & 0.02460(12) & 0.01403(12) & 0.00954(12) & 0.00649(12) \\
V04 & 0.386429 & 0.1276(2) & 0.0074(2) & 0.0100(2) & 0.0061(2) & 0.0034(2) \\
V13 & 0.331386 & 0.0654(3) & 0.0028(3) & 0.0008(3) & 0.0004(3) & -- \\
V14 & 0.317083 & 0.0362(4) & 0.0012(5) & -- & -- & -- \\
V16 & 0.338646 & 0.0363(4) & 0.0014(4) & -- & -- & -- \\
V17 & 0.415666 & 0.0228(5) & -- & -- & -- & -- \\
V18 & 0.430557 & 0.02653(13) & -- & -- & -- & -- \\
M80 15470 & 0.315231 & 0.0193(3) & 0.0006(3) & -- & -- & -- \\
\hline
\multicolumn{7}{c}{foreground stars}\\ 
V10 & 0.614147 & 0.1056(3) & 0.0579(3) & 0.0362(3) & 0.0263(3) & 0.0158(3) \\
V15 & 0.348275 & 0.0448(2) & 0.0042(2) & 0.0026(2) & 0.0016(2) & 0.0010(2) \\
\hline
\end{tabular}
\end{table*}

\begin{table*}
\centering 
\caption{Fourier phases of the M80 RR Lyrae stars.}
\label{tab:phase}
\begin{tabular}{lccccccc}  
\hline\hline
ID & P   & $\phi_1$ & $\phi_2$ & $\phi_3$ & $\phi_4$ & $\phi_5$  \\
 ~ & (d) & (rad) & (rad) & (rad) & (rad) & (rad)   \\
\hline
\multicolumn{7}{c}{RRab stars}\\
V05 & 0.663933 & 2.790(3) & 1.912(5) & 1.218(8) & 0.710(15) & 0.02(3) \\
V08 & 0.661808 & 0.777(2) & 4.044(4) & 1.222(7) & 4.890(10) & 2.136(19) \\
V09 & 0.664178 & 4.063(2) & 4.357(3) & 4.875(5) & 5.543(8) & 6.095(18) \\
V19 & 0.596075 & 0.031(6) & 2.829(14) & 5.89(2) & 2.91(4) & 5.95(7) \\
V20 & 0.745276 & 0.031(9) & 2.83(2) & 5.89(3) & 2.91(8) & 5.95(16) \\
V21 & 0.815007 & 0.540(2) & 4.011(6) & 1.450(15) & 5.37(4) & 3.37(6) \\
\hline
\multicolumn{7}{c}{RRc stars}\\  
V03 & 0.345711 & 0.9105(9) & 5.167(5) & 3.085(8) & 1.089(14) & 4.871(19) \\
V04 & 0.386429 & 0.7976(14) & 5.14(3) & 3.624(14) & 1.16(3) & 5.06(4) \\
V13 & 0.331386 & 3.198(5) & 3.62(10) & 3.97(48) & 4.97(91) & -- \\
V14 & 0.317083 & 4.087(13) & 3.98(54) & -- & -- & -- \\
V16 & 0.338646 & 6.150(12) & 2.56(35) & -- & -- & -- \\
V17 & 0.415666 & 5.259(18) & -- & -- & -- & -- \\
V18 & 0.430557 & 1.918(5) & -- & -- & -- & -- \\
M80 15470 & 0.315231 & 3.933(13) & 5.13(69) & -- & -- & -- \\
\hline
\multicolumn{7}{c}{foreground stars}\\ 
V10 & 0.614147 & 4.415(3) & 4.860(5) & 5.567(10) & 0.038(11) & 0.85(2) \\
V15 & 0.348275 & 4.547(5) & 6.14(5) & 1.66(8) & 3.30(13) & 4.35(22) \\
\hline
\end{tabular}
\end{table*}

\begin{table*}
\centering \small 
\caption{Fourier parameters and photometric metallicity estimates (JK/M: \citet{JK1996} and \citet{Morgan2007} relations; N13: \citet{nemec-2013} relations) of the M80 RR Lyrae stars.}
\label{tab:four}
\begin{tabular}{lccccccccccc}  
\hline\hline
ID & P   & R$_{21}$ & R$_{31}$ & $\phi_{21}$ & $\phi_{31}$ & $\sigma$R$_{21}$ & $\sigma$R$_{31}$ & $\sigma\phi_{21}$ & $\sigma\phi_{31}$ & [Fe/H]$_{\rm JK/M}$ & [Fe/H]$_{\rm N13}$   \\
 ~ & (d) & -- & -- & (rad) & (rad) & -- & -- & (rad) & (rad) & -- & --   \\
\hline
\multicolumn{12}{c}{RRab stars}\\
V05 & 0.663933 & 0.51363 & 0.32144 & 2.6149 & 5.4142 & 0.00086 & 0.00084 & 0.0029 & 0.0065 & -1.59 & -1.54  \\
V08 & 0.661808 & 0.54323 & 0.33749 & 2.4901 & 5.1742 & 0.00068 & 0.00071 & 0.003 & 0.0061 & -2.17 & -1.85   \\
V09 & 0.664178 & 0.53422 & 0.34116 & 2.5143 & 5.2516 & 0.00059 & 0.00058 & 0.0009 & 0.0011 & -1.98 & -1.76  \\
V19 & 0.596075 & 0.48174 & 0.27823 & 2.31 & 4.986 & 0.00072 & 0.00081 & 0.02 & 0.02 & -2.11 & -1.75   \\
V20 & 0.745276 & 0.41906 & 0.25677 & 2.768 & 5.799 & 0.00173 & 0.00164 & 0.288 & 0.288 & -1.10 & -1.46   \\
V21 & 0.815007 & 0.41325 & 0.19156 & 2.9313 & 6.1133 & 0.00024 & 0.00024 & 0.0046 & 0.0112 & -0.58 & -1.41  \\
\hline                                                                     
\multicolumn{12}{c}{RRc stars}\\
V03 & 0.345711 & 0.14646 & 0.08354 & 3.3458 & 0.3539 & 0.00017 & 0.00017 & 0.0014 & 0.0029 & -1.70 & -1.74  \\
V04 & 0.386429 & 0.05828 & 0.07868 & 3.5462 & 1.2309 & 0.00025 & 0.00026 & 0.0052 & 0.0043 & -1.62 & -1.66  \\
V13 & 0.331386 & 0.04312 & 0.01211 & 3.509 & 0.662 & 0.00043 & 0.0004 & 0.027 & 0.12 & -1.34 & -1.44  \\
V14 & 0.317083 & 0.03416 & <0.0076 & 2.087  & --& 0.0007  & --& 0.135  & -- & --& - \\
V16 & 0.338646 & 0.03714 & <0.0021 & 2.83  & --& 0.00055  & --& 0.138  & -- & --& - \\
V17 & 0.415666 & <0.1064 & <0.0611  & -- & -- & -- & -- & -- & -- & -- & --\\
V18 & 0.430397 & <0.0202 & <0.0242  & -- & -- & -- & -- & -- & -- & -- & --\\
M80 15470 & 0.315231 & 0.02949 & <0.0135 & 3.545  & --& 0.00035  & --& 0.134  & -- & -- & -  \\
\hline                                                                     
\multicolumn{12}{c}{foreground   stars}\\                
V10 & 0.614147 & 0.54815 & 0.34322 & 2.3145 & 4.8895 & 0.00044 & 0.00044 & 0.0012 & 0.0018 & -2.53 & -1.98  \\
V15 & 0.348275 & 0.09278 & 0.05787 & 3.3284 & 0.583 & 0.00031 & 0.00033 & 0.0084 & 0.049 & -1.69 & -1.66 \\
\hline
\end{tabular}
\end{table*}

\begin{table*}
\centering 
\caption{\textit{Gaia} absolute magnitudes, dereddened colors and zeropoint-corrected parallaxes.}
\label{tab:absmag}
\begin{tabular}{lccccccc}  
\hline
\hline
ID & Gaia DR3 ID & $M_{\rm G}$ & (BP--RP)$_0$ & $\sigma M_{\rm G}$ & $\sigma$(BP--RP)$_0$ & $\varpi$ &  $\sigma\varpi$ \\
~  & ~           & (mag)      & (mag)        & (mag)              & (mag)                & (mas)    &  (mas)          \\
\hline
V01 & 6050423169797108864 & -2.425 & 0.891 & 0.036 & 0.082 & 0.168 & 0.018 \\
V02 & 6050422104646047104 & -3.030 & 1.560 & 0.032 & 0.010 & 0.086 & 0.018 \\
$[$CBG2015$]$ 16162 & 6050422207724345216 & -3.189 & 1.632 & 0.032 & 0.016 & 0.127 & 0.019 \\
\hline
\multicolumn{8}{c}{RRab stars}\\
V05 & 6050422070285372032 & 0.372 & 0.596 & 0.034 & 0.051 & -0.035 & 0.050 \\
V08 & 6050422860560346496 & 0.337 & 0.620 & 0.034 & 0.057 & 0.084 & 0.061 \\
v09 & 6050422963639595648 & 0.355 & 0.596 & 0.034 & 0.040 & - & - \\
V10 & 6050422860560309888 & -0.711 & 0.758 & 0.032 & 0.021 & - & - \\
V19 & 6050422860560346240 & 0.384 & - & 0.035 & - & - & - \\
V20 & 6050422860560727936 & 0.310 & 0.721 & 0.034 & 0.019 & - & - \\
V21 & 6050422860561693696 & 0.258 & - & 0.034 & - \\
\hline
\multicolumn{8}{c}{RRc stars}\\
V03 & 6050422242084090752 & 0.300 & 0.334 & 0.033 & 0.038 & 0.251 & 0.050 \\
V04 & 6050423131140706432 & 0.347 & 0.432 & 0.032 & 0.030 & 0.169 & 0.049 \\
V13 & 6050422173364568576 & 0.375 & 0.357 & 0.032 & 0.014 & -0.096 & 0.052 \\
V14 & 6050423096777893248 & 0.364 & 0.337 & 0.032 & 0.015 & 0.000 & 0.049 \\
V15 & 6050422104646094976 & -0.737 & 0.882 & 0.032 & 0.014 & - & - \\
V16 & 6050423066717872128 & 0.310 & 0.346 & 0.032 & 0.010 & 0.048 & 0.046 \\
V17 & 6050422860560290432 & 0.343 & - & 0.034 & - & 0.141 & 0.061 \\
V18 & 6050422894920226944 & 0.409 & 0.533 & 0.032 & 0.008 & 0.118 & 0.048 \\
M80--15740 & 6050421967206133248 & 0.376 & 0.302 & 0.032 & 0.005 & - & - \\
 \hline
\multicolumn{8}{c}{SX Phe stars}\\
V22 & 6050422207724674816 & 2.886 & 0.526 & 0.033 & 0.184 & 2.51 & 0.31 \\
V23 & 6050422929272927744 & 3.023 & 0.407 & 0.032 & 0.153 & -0.99 & 0.30 \\
V24 & 6050422070286297728 & 2.830 & - & 0.043 & - & - & - \\
\hline
\multicolumn{8}{c}{other variables}\\
V25 & 6050423238510697216 & 4.132 & 0.628 & 0.032 & 0.103 & 0.88 & 0.46 \\
V27 & 6050422860560452480 & 1.971 & - & 0.035 & - & - & - \\
V28 & 6050421619312200576 & 3.589 & -0.294 & 0.032 & 0.056 & -0.44 & 0.33 \\
V29 & 6050422207725742848 & 1.801 & 0.906 & 0.033 & 0.040 & - & - \\
EPIC 204270267 & 6050421928546399360 & 3.056 & 1.275 & 0.032 & 0.053 & 0.34 & 0.25 \\
\hline
\end{tabular}
\end{table*}

\end{appendix}

\end{document}